\documentclass[a4paper, 11pt, sort&compress]{elsarticle}

\usepackage{float}
\usepackage{multirow}
\usepackage{subfigure}
\usepackage{amsmath}
\usepackage{mathtools}
\usepackage{subfiles}
\usepackage{multirow}
\usepackage{multicol}
\usepackage[english]{babel}
\usepackage{blindtext}
\usepackage{algorithm}
\usepackage{algpseudocode}
\usepackage{amssymb}
\usepackage{bbm, bbold}
\usepackage{xcolor}
\usepackage{bbm, bbold}
\usepackage{bm}
\usepackage{hyperref}
\usepackage{array}

\newcommand{\pderiv}[2]{\frac{\partial #1}{\partial #2}}

\newcommand{\bs}[1]{\boldsymbol{#1}}

\newcommand{\dOmega}{\; \mathrm{d}\Omega}
\newcommand{\dGamma}{\; \mathrm{d}\Gamma}

\newcommand{\dOmegae}{\; \mathrm{d}\Omega^e}

\RequirePackage[margin=20mm]{geometry}

\begin{document}

\fontfamily{ptm}\selectfont

\begin{frontmatter}

\title{Accurate iteration-free mixed-stabilised formulation for laminar incompressible Navier-Stokes: Applications to fluid-structure interaction}

\author[add1]{Chennakesava Kadapa\corref{cor1}}
\ead{c.kadapa@swansea.ac.uk}

\author[add2]{Wulf G Dettmer}

\author[add2]{Djordje Peri\'c}

\cortext[cor1]{Corresponding author}

\address[add1]{Swansea Academy of Advanced Computing, Swansea University, SA1 8EN, United Kingdom}
\address[add2]{Zienkiewicz Centre for Computational Engineering (ZCCE), Swansea University, SA1 8EN, United Kingdom}

\begin{abstract}
Stabilised mixed velocity-pressure formulations are one of the widely-used finite element schemes for computing the numerical solutions of laminar incompressible Navier-Stokes. In these formulations, the Newton-Raphson scheme is employed to solve the nonlinearity in the convection term. One fundamental issue with this approach is the computational cost incurred in the Newton-Raphson iterations at every load/time step. In this paper, we present an iteration-free mixed finite element formulation for incompressible Navier-Stokes that preserves second-order temporal accuracy of the generalised-alpha and related schemes for both velocity and pressure fields. First, we demonstrate the second-order temporal accuracy using numerical convergence studies for an example with a manufactured solution. Later, we assess the accuracy and the computational benefits of the proposed scheme by studying the benchmark example of flow past a fixed circular cylinder. Towards showcasing the applicability of the proposed technique in a wider context, the inf-sup stable P2-P1 pair for the formulation without stabilisation is also considered. Finally, the resulting benefits of using the proposed scheme for fluid-structure interaction problems is illustrated using two benchmark examples in fluid-flexible structure interaction.
\end{abstract}


\end{frontmatter}

\section{Introduction} \label{section-intro}
Navier-Stokes equations which model fluid flow are one of the most widely used partial differential equations in the fields of science and engineering. Obtaining accurate numerical solutions of Navier-Stokes equations has been one of the frontiers of research for more than a century. With the advent of computers, development of accurate, robust and computationally efficient schemes for computing the numerical solutions of Navier-Stokes equations has been one of the principal areas of research in computational engineering and numerical mathematics.

The vast majority of schemes available for numerical solutions of Navier-Stokes belong to either the finite difference method (FDM) or finite volume method (FVM) or finite element method (FEM). Finite volume methods have been the front runners of computational fluid dynamics (CFD) and still enjoy a significant share among the commercial and opensource software suites for CFD, for example, OpenFOAM and ANSYS Fluent. Thanks to the research and developments during the past couple of decades, see \cite{book-cfd-DoneaHuerta, book-cfd-Elman, book-cfd-Layton, book-cfd-Zienkiewicz, book-fem-GreshoSaniVol2} and references therein, finite element methods are not only proving to be strong competitors to finite volume methods but also are surpassing them in simulating multiphysics problems, see \cite{book-fsi-Bazilevs, book-fsi-Bungartz, book-fsi-Bungartz2}; several recent open-source and commercial software tools, for example, COMSOL Multiphysics, deal.ii, oomph-lib, FEBio, Elmer FEM, and FEniCS, are based on FEM. The increased interest in the finite element methods for applications in CFD is their advantages in modelling complex multiphysics phenomenon such as fluid-structure interaction (FSI).

Among finite element methods for incompressible fluid flow problems, those based on the SUPG/PSPG stabilisation have been widely explored in the literature, thanks to their computational advantages, especially in using the equal-order interpolation for velocity and pressure fields and iterative solvers that are computationally efficient for large matrix systems resulting from three-dimensional problems. Seminal contributions to stabilised mixed methods for incompressible fluid flow problems are by Hughes, Franca and group \cite{BrooksCMAME1982, HughesFrancaCMAME1986} and Tezduyar et al. \cite{TezduyarAAM1991, TezduyarCMAME1992, TezduyarCMAME2000}.

Stabilised mixed formulations for incompressible Navier-Stokes have been successfully adapted to the multiphysics problem of fluid-structure interaction by several researchers. For comprehensive details on fluid-structure interaction using mixed-stabilised methods, we refer the reader to Bazilevs et al. \cite{book-fsi-Bazilevs}, Bungartz and Sch\"afer \cite{book-fsi-Bungartz}, and Bungartz et al. \cite{book-fsi-Bungartz2}, and references therein. In the majority of the FSI problems, the fluid solver is the most time-consuming part; therefore, for efficient simulation of FSI problems, it is always advantageous to minimise the computational cost incurred in the fluid solver, which is the main motivation behind the proposed work.

The source of nonlinearity in the laminar viscous incompressible Navier-Stokes equations is the convection term, and in the case of mixed-stabilised finite element formulations for incompressible Navier-Stokes, stabilisation terms also contribute to nonlinearity, see \cite{BrooksCMAME1982, TezduyarAAM1991}. In the mixed-stabilised formulations, this nonlinearity is treated by employing an iterative solution scheme, for example, the widely-used Newton-Raphson scheme. However, such iterative schemes have several disadvantages, with the computational cost being the most important one. With the Newton-Raphson scheme, one must re-compute the Jacobian matrix and solve the matrix system at every iteration of every load/time step. Even though modified Newton-Raphson schemes that do not require computation of tangent matrix during every iteration are available, their convergence is poor at best; in some cases, they may not converge at all. Because of these disadvantages, it is always beneficial to avoid using such iterative solution schemes.

Some variants of projection and fractional-step schemes c.f. \cite{book-cfd-Ferziger, book-cfd-Zienkiewicz} avoid this nonlinearity with a linear approximation of the nonlinear convection term. However, the fundamental issue with such an approach is their first-order temporal convergence which poses severe restrictions on time step sizes in order to obtain accurate numerical results. Although high-order temporal accuracy can be achieved by employing extrapolated velocity fields, such a technique of using solutions from previous time steps is not always feasible, especially for cut-cell based formulations in which the fluid domain changes over time c.f. \cite{DettmerCMAME2016, KadapaCMAME2017rigid, KadapaCMAME2018}. To the knowledge of the authors, no literature exists on the mixed-stabilised formulations that employ linearised convection operators for laminar incompressible Navier-Strokes. In this paper, we present a linearised version of mixed velocity-pressure formulation with SUPG/PSPG stabilisation for laminar incompressible Navier-Stokes that preserves second-order accuracy of the generalised-alpha scheme, and that can compute numerical results that are comparable with those of the standard iteration-based scheme while requiring only a fraction of computational resources. We also demonstrate the applicability of the proposed technique in a wider context by using the inf-sup stable P2-P1 pair for the formulation without stabilisation.

The rest of the paper is organised as follows. The governing equations, the finite element formulation and the time integration schemes used in the present work are detailed in Section \ref{section-formulation}. The accuracy and the computational benefits of the proposed scheme are first demonstrated using flow over fixed body-fitted meshes in Section \ref{section-examples-fixed}. Later, the applicability, accuracy and the resulting computational benefits of using the proposed approach for fluid-structure interaction problems is illustrated using two fluid-flexible structure interaction benchmark examples in Section \ref{section-examples-fsi}. Important contributions and observations of the present work are summarised, and conclusions are drawn in Section \ref{section-conclusion}.

\section{Formulation} \label{section-formulation}
\subsection{Governing equations}
The initial-boundary value problem for a laminar, isothermal, viscous, incompressible fluid flow in the domain $\Omega$ is stated as: \\
Find velocity, $\bm{v}:\Omega \rightarrow \mathbb{R}^3$; and pressure, $p: \Omega \rightarrow \mathbb{R}$, such that:
\begin{subequations} \label{NS-main}
\begin{align}
\rho \, \dot{\bm{v}} + \rho \, (\bm{v} \cdot \nabla) \, \bm{v} - \mu \, \nabla^2 \bm{v} + \nabla p &= \bm{g} \hspace*{20mm} \mathrm{in} \quad \Omega  \label{NS-main-eqn1} \\
 \nabla \cdot \bm{v} &= 0 \hspace*{22mm} \mathrm{in} \quad \Omega \\
 \bm{v} &= \bar{\bm{v}} \hspace*{20mm} \mathrm{on} \quad \Gamma_\mathrm{D} \\
 \bs{\sigma} \cdot \bm{n} &= \bar{\bm{t}} \hspace*{21mm} \mathrm{on} \quad \Gamma_\mathrm{N} \\
 \bm{v}(\cdot,0) &= \bm{v}_0  \hspace*{20mm} \mathrm{in} \quad \Omega \\
 p(\cdot,0) &= p_0  \hspace*{20mm} \mathrm{in} \quad \Omega
\end{align}
\end{subequations}
where $\dot{\bm{v}}=\pderiv{\bm{v}}{t}$; $\nabla$ is the gradient operator; $\rho$ is the density of the fluid; $\mu$ is the dynamic viscosity of the fluid; $\bm{g}$ is the body force; $\bm{n}$ is the unit outward normal on the boundary, $\Gamma$, of $\Omega$; $\Gamma_\mathrm{D}$ is the part of the boundary where Dirichlet boundary condition $\bar{\bm{v}}$ is applied; and $\Gamma_\mathrm{N}$ is the part of the boundary where Neumann boundary condition $\bar{\bm{t}}$ is applied. Here, $\Gamma = \Gamma_\mathrm{D} \cup \Gamma_\mathrm{N}$ and $\Gamma_\mathrm{D} \cap \Gamma_\mathrm{N} = \emptyset$. $\bm{v}_0$ and $p_0$ are the initial velocity and pressure fields in the fluid in the domain. The pseudo-stress $\bs{\sigma}$ is given by,
\begin{align} \label{stress-ss}
\bs{\sigma} = \mu \nabla \bm{v} - p \, \bm{I}
\end{align}
where $\bm{I}$ is the second-order identity tensor.

\subsection{Variational statement}
The motivation behind the proposed technique is to improve the computational efficiency of coupled fluid-structure interaction simulations using the stabilised finite element formulations for fluid flow \cite{DettmerArchCME2007, DettmerCMAME2016, KadapaCMAME2017rigid, KadapaCMAME2018}. Nevertheless, we also demonstrate that the proposed technique is equally applicable for the mixed formulation with inf-sup stable or Ladyzhenskaya-Babuska-Brezzi (LBB) stable elements; the classical Taylor-Hood elements based on the Langrangian polynomials \cite{book-fem-BrezziFortin, book-cfd-Elman} are employed. In the computer implementation with the inf-sup stable pair, the stabilisation terms are dropped.

The weak statement for the incompressible Navier-Stokes with velocity and pressure as the primary solution variables is: \\
Find the discretised fluid velocity, $\bm{v}^h \in \mathcal{V}^h$, and discretised pressure, $p^h \in \mathcal{P}^h$, such that for all the weighting functions $\bm{w}^h \in \mathcal{W}^h$ and $q^h \in \mathcal{Q}^h$
\begin{align} \label{weakform-eqn1}
\mathcal{I}_{1} + \mathcal{I}_{2} = \int_{\Omega} \bm{w}^h \cdot \bm{g}_{\phi_2} \dOmega +  \int_{\Gamma_{\mathrm{N}}} \bm{w}^h \cdot \bar{\bm{t}}_{\phi_2} \dGamma
\end{align}
where $\mathcal{V}^h$ and $\mathcal{W}^h$ are the function and test spaces for the velocity field; $\mathcal{P}^h$ and $\mathcal{Q}^h$ are the function and test spaces for the pressure field; $\mathcal{I}_{1}$ is the collection of linear terms in the weak form, given by,
\begin{align} \label{weakform-I1}
\mathcal{I}_{1} = \int_{\Omega} \rho \, \bm{w}^h \cdot \dot{\bm{v}}^h_{\phi_1}  \dOmega
+ \int_{\Omega} \mu \, \nabla \bm{w}^h : \nabla \bm{v}^h_{\phi_2} \dOmega
- \int_{\Omega} (\nabla \cdot \bm{w}^h) \, p^h_{\phi_2} \, d \Omega + \int_{\Omega} q^h \, (\nabla \cdot \bm{v}^h_{\phi_2}) \dOmega
\end{align}
and $\mathcal{I}_{2}$ is the collection of nonlinear terms, given by,
\begin{align} \label{weakform-I2}
\mathcal{I}_{2} 
&= \int_{\Omega} \rho \, \bm{w}^h \cdot \left( \bm{v}^h_{\phi_3} \cdot \nabla \bm{v}^h_{\phi_2} \right) \dOmega + \sum_{e=1}^{\mathsf{nelem}} \int_{\Omega^{e}} \frac{1}{\rho} [ \tau_{\text{SUPG}} \, \rho \, \bm{v}^h_{\phi_4} \cdot \nabla \bm{w}^h + \tau_{\text{PSPG}} \, \nabla q^h ] \cdot \bm{r}_{\text{M}} \dOmegae
\end{align}
where $\sum_{e=1}^{\mathsf{nelem}}$ is the summation over the elements, $\mathsf{nelem}$ is the number of elements, $\tau_{\text{SUPG}}$ and $\tau_{\text{PSPG}}$ are the stabilisation parameters, and $\bm{r}_{\text{M}}$ is the residual of the momentum equation, given as,
\begin{align} \label{weakform-resi1}
\bm{r}_{\text{M}} = \rho \, \dot{\bm{v}}^h_{\phi_1} + \rho ( \bm{v}^h_{\phi_3} \cdot \nabla \bm{v}^h_{\phi_2} ) - \mu \nabla^2 \bm{v}^h_{\phi_2} + \nabla p^h_{\phi_2}.
\end{align}
In the above equations (\ref{weakform-eqn1})-(\ref{weakform-resi1}), $\phi_1$, $\phi_2$, $\phi_3$ and $\phi_4$ are the time instants which are chosen depending on the time integration scheme and the treatment of convection term as a nonlinear or a linearised one.

\subsection{Time integration schemes}
We consider three different time integration schemes: BDF1, BDF2 and generalised-alpha.
\subsubsection{BDF schemes}
With $\bm{v}^{h}_{n+1}$, $\bm{v}^{h}_{n}$ and $\bm{v}^{h}_{n-1}$ as the velocity of the fluid at time instants $t_{n+1}$, $t_{n}$ and $t_{n-1}$, respectively, the acceleration of the fluid $\dot{\bm{v}}^{h}_{n+1}$ at time instant $t_{n+1}$, for the first two BDF schemes is given as
\begin{align}
\textrm{BDF1:} && 
\dot{\bm{v}}^{h}_{n+1} &= \frac{1}{\Delta t} \; \left[ \bm{v}^{h}_{n+1} - \bm{v}^{h}_{n} \right],  \label{BDF1-defs1} \\
\textrm{BDF2:} && 
\dot{\bm{v}}^{h}_{n+1} &= \frac{1}{2 \, \Delta t} \; \left[3 \, \bm{v}^{h}_{n+1} - 4 \, \bm{v}^{h}_{n} + \bm{v}^{h}_{n-1} \right],  \label{BDF2-defs1}
\end{align}
where $\Delta t$ is the time step size. BDF1 scheme is popular as the backward Euler (BE) scheme. The order of temporal accuracy of the BDF1 and BDF2 schemes is, respectively, one and two.

\subsubsection{Generalised-alpha time integration scheme}
For the generalised-alpha scheme proposed by Jansen et al. \cite{JansenCMAME2000}, and referred to as \textit{GA scheme} henceforth in this paper, the relations between velocity and acceleration are given as
\begin{align}
\bm{v}^{h}_{n+\alpha_f} &= \alpha_f \, \bm{v}^{h}_{n+1} + (1-\alpha_f) \, \bm{v}^{h}_{n},  \label{GA-fluid-defs1} \\
\dot{\bm{v}}^{h}_{n+\alpha_m} &=  \alpha_m \, \dot{\bm{v}}^{h}_{n+1} + (1-\alpha_m) \, \dot{\bm{v}}^{h}_{n},  \label{GA-fluid-defs2} \\
\dot{\bm{v}}^{h}_{n+1} &= \frac{1}{\gamma \Delta t} \; (\bm{v}^{h}_{n+1} - \bm{v}^{h}_{n}) + \frac{\gamma-1}{\gamma} \; \dot{\bm{v}}^{h}_{n}. \label{GA-fluid-defs3}
\end{align}
where $\dot{\bm{v}}^{h}_n$ is the acceleration of the fluid at time instant $t_n$. Following \cite{JansenCMAME2000, KadapaCS2017}, the GA scheme is second-order accurate if the parameters $\alpha_m$, $\alpha_f$ and $\gamma$ are chosen such that
\begin{align} \label{GA-fluid-params}
\alpha_m = \frac{1}{2} \; \frac{3 - \rho_{\infty}}{1+\rho_{\infty}}, \quad
\alpha_f = \frac{1}{1 + \rho_{\infty}}, \quad
\gamma   = \frac{1}{2} + \alpha_m - \alpha_f, \quad \mathrm{for} \quad 0 \leq \rho_\infty \leq 1.
\end{align}
Here, the parameter $\rho_\infty$ is the maximum absolute eigenvalue of the amplification matrix (also referred as the spectral radius) of the GA scheme as $\Delta t \rightarrow \infty$. This single parameter gives the user the control over the amount of numerical damping for a particular resolution. By adjusting $\rho_{\infty}$, the user can control the range of frequencies, relative to the chosen resolution level, that are to be preserved or damped out. For $\rho_\infty=1$, the scheme possesses zero numerical damping, meaning that all the modes are preserved. Spectrally, this case is equivalent to the trapezoidal scheme. For $\rho_\infty=0$, the scheme annihilates all the high-frequency modes after the first time step. The GA scheme is spectrally equivalent to the second-order accurate backward-difference formula (BDF2), for $\rho_\infty=0$, see \cite{LovricCMAME2018} for details. Thus, by choosing $\rho_{\infty}$ appropriately, different time integration schemes can be recovered from the GA scheme.

\subsection{Formulation with nonlinear convection operator} \label{subsection-formulation-nonlin}
The conventional method of solving incompressible Navier-Stokes using the mixed-stabilised finite element formulations is to use the Newton-Raphson scheme to resolve the nonlinearity in the convection terms in equations (\ref{weakform-I2}) and (\ref{weakform-resi1}). For the purpose of discussion throughout this paper, we refer the conventional method of using the Newton-Raphson scheme as the \emph{standard} scheme.

For the standard scheme, the time instants $\phi_1$, $\phi_2$, $\phi_3$ and $\phi_4$ in equations (\ref{weakform-eqn1})-(\ref{weakform-resi1}) for the BDF schemes become
\begin{align}
\phi_1  &= \phi_2 = \phi_3 = \phi_4 = n+1,
\end{align}
while the corresponding values for the GA scheme are
\begin{align}
\phi_1  &= n+\alpha_m  \\
\phi_2  &= \phi_3 = \phi_4 = n+\alpha_f.
\end{align}

Following the above discussion, the generalised form of the functional $\mathcal{I}_2$, for the standard scheme can be written as,
\begin{align} \label{weakform-FI}
\mathcal{I}_{\text{2-Standard}}
&= \int_{\Omega} \rho \, \bm{w}^h \cdot \left( \bm{v}^h_{n+\alpha_f} \cdot \nabla \bm{v}^h_{n+\alpha_f} \right) \dOmega \nonumber  \\
&+ \sum_{e=1}^{nel} \int_{\Omega^{e}} \frac{1}{\rho} [ \tau_{\text{SUPG}} \, \rho \, \bm{v}^h_{n+\alpha_f} \cdot \nabla \bm{w}^h + \tau_{\text{PSPG}} \, \nabla q^h ] \cdot \bm{r}_{\text{M-Standard}} \dOmegae,
\end{align}
where,
\begin{align}  \label{resi-FI}
\bm{r}_{\text{M-Standard}}  &= \rho \, \dot{\bm{v}}^h_{n+\alpha_m} - \mu \nabla^2 \bm{v}^h_{n+\alpha_f} + \nabla p_{n+\alpha_f} + \rho ( \bm{v}^h_{n+\alpha_f} \cdot \nabla \bm{v}^h_{n+\alpha_f} ),
\end{align}
which can be adapted for the BDF schemes by setting $\alpha_m=1$ and $\alpha_f=1$. 

In spite of its second-order convergence of residuals with respect to iterations, the Newton-Raphson scheme employed to resolve the nonlinearities in the standard Galerkin and stabilisation terms in equation (\ref{weakform-I2}) becomes quite expensive, especially for large-scale problems in three dimensions. Moreover, our experience indicates that such an iterative procedure is unnecessary for all practical applications involving laminar fluid flows, particularly when small enough time steps are used for fluid-structure interaction problems due to accuracy and stability concerns. In the following subsection, we present a novel iteration-free technique for computing numerical solutions using mixed-stabilised formulations for unsteady laminar fluid flow problems without compromising the temporal accuracy for first- and, more importantly, second-order accurate time integration schemes.

\subsection{Formulation with linearised convection operators} \label{subsection-formulation-lin}
The motivation behind the proposed scheme is to avoid iteration solution techniques at every time step while preserving the accuracy for the finite element schemes for incompressible Navier-Stokes equations when employing the coupled velocity-pressure mixed formulation, either with a stabilised formulation or with the LBB-stable elements.

Considering that $\phi_4=n$, the only nonlinearity present in the whole functional $\mathcal{I}_2$ is the convection term, $\rho \, \left(\bm{v}^h_{\phi_3} \cdot \nabla \right) \bm{v}^h_{\phi_2}$. Iterative solution techniques in the fractional-step and projection schemes are avoided by choosing $\bm{v}^h_{\phi_3}=\bm{v}^h_{n}$, which, unfortunately, limits the order of temporal accuracy to one, even with higher-order accurate time integration schemes for the viscous and pressure gradient terms. Second-order temporal accuracy can be achieved by choosing appropriate extrapolation for the convective velocity field, see \cite{LovricCMAME2018} for the details.

The generalised form of the functional $\mathcal{I}_2$ for the linearised convection operator with the first- and second-order accurate extrapolated convection velocity can now be written as
\begin{align} \label{weakform-extrp}
\mathcal{I}_{\text{2-Extrapolated}}
&= \int_{\Omega} \rho \, \bm{w}^h \cdot \left( \widetilde{\bm{v}}^h_{n+\alpha_f} \cdot \nabla \bm{v}^h_{n+\alpha_f} \right) \dOmega \nonumber  \\
&+ \sum_{e=1}^{nel} \int_{\Omega^{e}} \frac{1}{\rho} [ \tau_{\text{SUPG}} \, \rho \, \bm{v}^h_{n} \cdot \nabla \bm{w}^h + \tau_{\text{PSPG}} \, \nabla q^h ] \cdot \bm{r}_{\text{M-Extrapolated}} \dOmegae
\end{align}
where,
\begin{align} \label{resi-extrp}
\bm{r}_{\text{M-Extrapolated}}  &= \rho \, \dot{\bm{v}}^h_{n+\alpha_m} - \mu \nabla^2 \bm{v}^h_{n+\alpha_f} + \nabla p_{n+\alpha_f} + \rho ( \widetilde{\bm{v}}^h_{n+\alpha_f} \cdot \nabla \bm{v}^h_{n+\alpha_f} ).
\end{align}
Here, $\widetilde{\bm{v}}^{h}_{n+\alpha_f}$ is the generalised extrapolated convection velocity which is necessary for preserving the second-order accuracy for the pressure field when using the GA scheme with $\rho_{\infty}>0$, and it is given as
\begin{align}
\widetilde{\bm{v}}^{h}_{n+\alpha_f} = \alpha_f \, \bm{v}^{h}_{*} + (1-\alpha_f) \, \bm{v}^{h}_{n}.
\end{align}
where
\begin{align}
\bm{v}^{h}_{*} &= \bm{v}^{h}_{n}, && \text{first-order accurate},  \label{eqn-extrap-velo-1} \\
\bm{v}^{h}_{*} &= 2 \, \bm{v}^{h}_{n} - \bm{v}^{h}_{n-1},  && \text{second-order accurate}. \label{eqn-extrap-velo-2}
\end{align}
In this work, the mixed-stabilised formulations with linearised convection operators with extrapolated convection velocity are denoted as \emph{Extrapolated1} and \emph{Extrapolated2}, respectively, with the first-order accurate (\ref{eqn-extrap-velo-1}) and second-order accurate (\ref{eqn-extrap-velo-2}) extrapolated convection velocity.

Although the mixed-stabilised formulation with the functional (\ref{weakform-extrp}) is completely linear, such a formulation involving velocity fields from the previous time steps, especially at $t_{n-1}$, is not feasible for schemes in which the active solution domain for the fluid problem changes over time, for example, schemes based on cut cells \cite{DettmerCMAME2016, KadapaCMAME2017rigid, KadapaCMAME2018}. Therefore, alternative strategies for linearising the convection operator while still preserving second-order temporal accuracy is necessary. In this work, the nonlinear functional (\ref{weakform-I2}) is reduced to a linear one as follows.

For $\rho=1$, the convection term for the GA scheme becomes
\begin{align} \label{eqn-convection1}
\bm{C} = \left(\bm{v}^h_{n+\alpha_f} \cdot \nabla \right) \bm{v}^h_{n+\alpha_f}.
\end{align}
With $\Delta \bm{v}^h$ as the velocity increment from time instant $t_n$ to time instant $t_{n+1}$, we have,
\begin{align} \label{eqn-velocity2}
\bm{v}^h_{n+\alpha_f} = \alpha_f \bm{v}^h_{n+1} + (1-\alpha_f) \bm{v}^h_n = \bm{v}^h_{n} + \alpha_f \Delta \bm{v}^h
\end{align}
Using (\ref{eqn-velocity2}), the convection term, $\bm{C}$, can be expanded as,
\begin{align} \label{eqn-convection2}
\bm{C} = \left(\bm{v}^h_{n+\alpha_f} \cdot \nabla \right) \bm{v}^h_{n+\alpha_f} 
&= \underbrace{\left( \bm{v}^h_n \cdot \nabla \right) \bm{v}^h_{n+\alpha_f}} + \underbrace{\left( \alpha_f \Delta \bm{v}^h \cdot \nabla \right) \bm{v}^h_{n+\alpha_f}} \\
&= \hspace*{12mm} \bm{C}_1 \hspace*{7mm} + \hspace*{14mm} \bm{C}_2 \nonumber
\end{align}
If we ignore the second term, $\bm{C}_2$, then the convection term becomes linear. However, as already discussed above, such a formulation is only first-order accurate in time. To overcome this drawback, we propose the following modification that helps in preserving the second-order accuracy of the GA scheme.

By rearranging $\bm{C}_2$, Eq. (\ref{eqn-convection2}) can be rewritten as,
\begin{align}
\left(\bm{v}^h_{n+\alpha_f} \cdot \nabla \right) \bm{v}^h_{n+\alpha_f} 
&= \left( \bm{v}^h_n \cdot \nabla \right) \bm{v}^h_{n+\alpha_f} + \left( \bm{v}^h_{n+\alpha_f} \cdot \nabla \right) \bm{v}^h_n - \left( \bm{v}^h_n \cdot \nabla \right) \bm{v}^h_n + \left( \alpha_f \Delta \bm{v}^h \cdot \nabla \right) \alpha_f \Delta \bm{v}^h \\
& \approx \left( \bm{v}^h_n \cdot \nabla \right) \bm{v}^h_{n+\alpha_f} + \left( \bm{v}^h_{n+\alpha_f} \cdot \nabla \right) \bm{v}^h_n - \left( \bm{v}^h_n \cdot \nabla \right) \bm{v}^h_n 
\end{align}
So, in the proposed scheme, we ignore only the higher-order (nonlinear) term $\left( \alpha_f \Delta \bm{v}^h \cdot \nabla \right) \alpha_f \Delta \bm{v}^h$, instead of ignoring the term $\bm{C}_2$ altogether. We demonstrate with convergence studies in Section. \ref{section-examples-fixed} that this approach preserves second-order temporal accuracy of the GA scheme for both the velocity and pressure fields.

Now, the functional $\mathcal{I}_2$, for the proposed scheme becomes,
\begin{align}  \label{weakform-SI}
\mathcal{I}_{\text{2-Proposed}}
&= \int_{\Omega} \rho \bm{w}^h \cdot \left( \bm{v}^h_{n} \cdot \nabla \bm{v}^h_{n+\alpha_f} + \bm{v}^h_{n+\alpha_f} \cdot \nabla \bm{v}^h_{n} - \bm{v}^h_{n} \cdot \nabla \bm{v}^h_{n} \right) \dOmega  \nonumber \\
&+ \sum_{e=1}^{nel} \int_{\Omega^{e}} \frac{1}{\rho} [ \tau_{\text{SUPG}} \, \rho \, \bm{v}^h_{n} \cdot \nabla \bm{w}^h + \tau_{\text{PSPG}} \, \nabla q^h ] \cdot \bm{r}_{\text{M-Proposed}} \dOmegae
\end{align}
where,
\begin{align}  \label{resi-SI}
\bm{r}_{\text{M-Proposed}} &= \rho \, \dot{\bm{v}}^h_{n+\alpha_m} - \mu \nabla^2 \bm{v}^h_{n+\alpha_f} + \nabla p_{n+\alpha_f} + \rho ( \bm{v}^h_{n} \cdot \nabla \bm{v}^h_{n+\alpha_f} + \bm{v}^h_{n+\alpha_f} \cdot \nabla \bm{v}^h_{n} - \bm{v}^h_{n} \cdot \nabla \bm{v}^h_{n} ).
\end{align}
The formulation for the BDF schemes can be recovered by setting $\alpha_m=1$ and $\alpha_f=1$ in the above expressions (\ref{weakform-SI}) and (\ref{resi-SI}).

Since all the higher-order (nonlinear) terms are eliminated, the proposed scheme is linear; therefore, it doesn't require an iterative approach. It can be implemented in a computer code to compute the solutions directly at the time instant $t_{n+1}$, in the same manner as that usually followed for the Stokes or Oseen equations.

It is also worth pointing out at this point that the proposed scheme does not require any special treatment to preserve the second-order accuracy in the case of adaptive time-stepping while the coefficients in equation (\ref{eqn-extrap-velo-2}) need appropriate modifications for the Extrapolated2 scheme. Note also that the computational cost incurred in computing and assembling the element matrices for one loop over elements is approximately the same in all the three schemes.

\graphicspath{{./figures/}}

\section{Numerical examples - flow over fixed meshes} \label{section-examples-fixed}
The accuracy and the computational advantages of the proposed technique are demonstrated first by studying flows over fixed body-fitted meshes. As the inf-sup stable pair, we use the P2-P1 element. For spatial discretisation when using the stabilised formulation, we employ the bi-linear quadrilateral (Q1) element with equal-order interpolation for velocity and pressure, which we denote as \emph{Q1-Stab} element for the sake of discussion. The stabilisation parameters \cite{TezduyarCMAME2000} are computed as,
\begin{align}
\tau_{\text{SUPG}} = \tau_{\text{PSPG}} &= \left[ \boldsymbol{v} \cdot \bm{G} \, \boldsymbol{v} + 4 \, \nu \, \bm{G}:\bm{G} \right]^{-1/2},
\end{align}
where $\nu=\frac{\mu}{\rho}$ and $\bm{G}$ is approximated as
\begin{align}
\bm{G} = \frac{4}{h^2} \, \boldsymbol{I},
\end{align}
in which $h=\sqrt{4 \, A/\pi}$ is the element characteristic length, with $A$ as the area of the element. The resulting matrix system of equations for the all the simulations presentted in this work are solved by using a sparse direct solver PARDISO \cite{pardiso}. The convergence tolerance for the residual in the standard iteration-based scheme is taken as $10^{-8}$.

Note that the formulations for the steady-state problems can be recovered by ignoring the time derivative terms and setting $\alpha_f=1$ in the expressions in Sections. \ref{subsection-formulation-nonlin} and \ref{subsection-formulation-lin}. For the sake of simplicity, only unsteady flow problems are considered in this work.

\subsection{Temporal convergence study with a manufactured solution}
\begin{figure}[H]
\centering
\includegraphics[trim = 0mm 0mm 0mm 0mm, clip, scale=0.65]{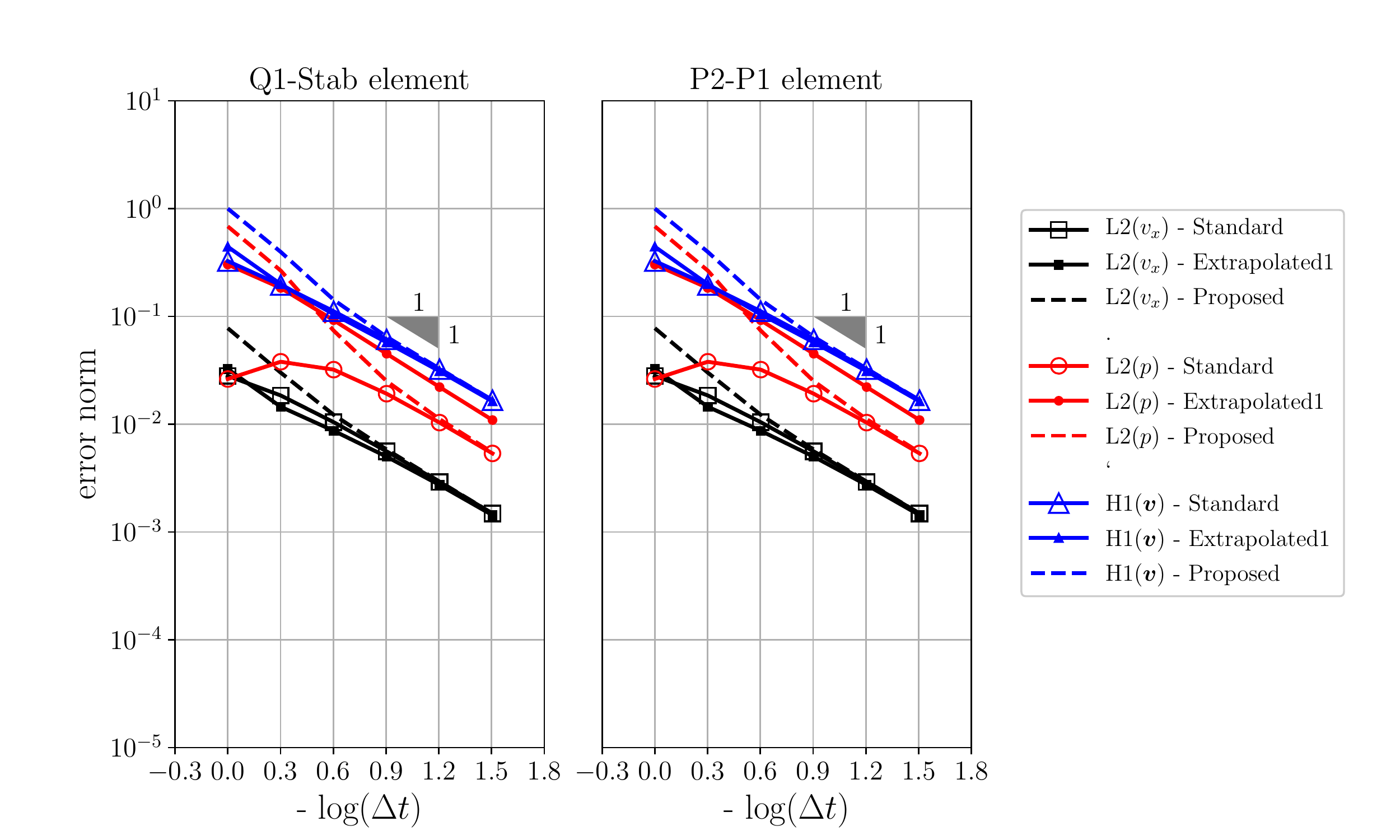}
\caption{Manufactured solution example: convergence of error norms at $t=5$ obtained with BDF1 scheme using different formulations. L2($\bullet$) and H1($\bullet$) refer to the L2 norm and H1 norm of the field $\bullet$, respectively. L2 norms in Y-component of velocity ($v_y$) are about the same as those of the X-component; therefore, are not included in the graph for the sake of clarity.}
\label{manfsoln-graphs-BE-quad}
\end{figure}

\begin{figure}[H]
\centering
\includegraphics[trim = 0mm 0mm 0mm 0mm, clip, scale=0.65]{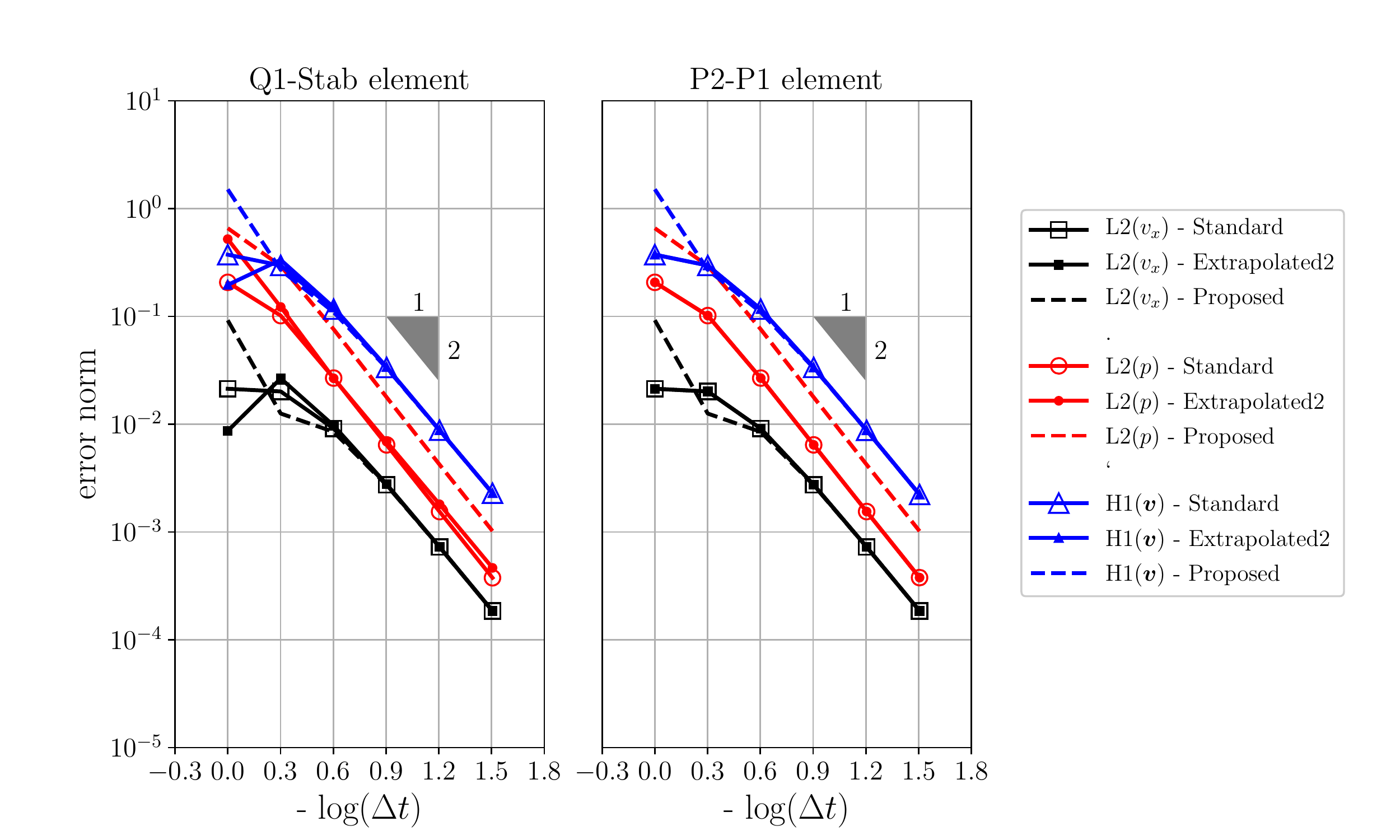}
\caption{Manufactured solution example: convergence of error norms at $t=5$ obtained with GA scheme with $\rho_{\infty}=0.0$ using different formulations. The error norm values for this case are identical to those of the BDF2 scheme.}
\label{manfsoln-graphs-GA1-quad}
\end{figure}

\begin{figure}[H]
\centering
\includegraphics[trim = 0mm 0mm 0mm 0mm, clip, scale=0.65]{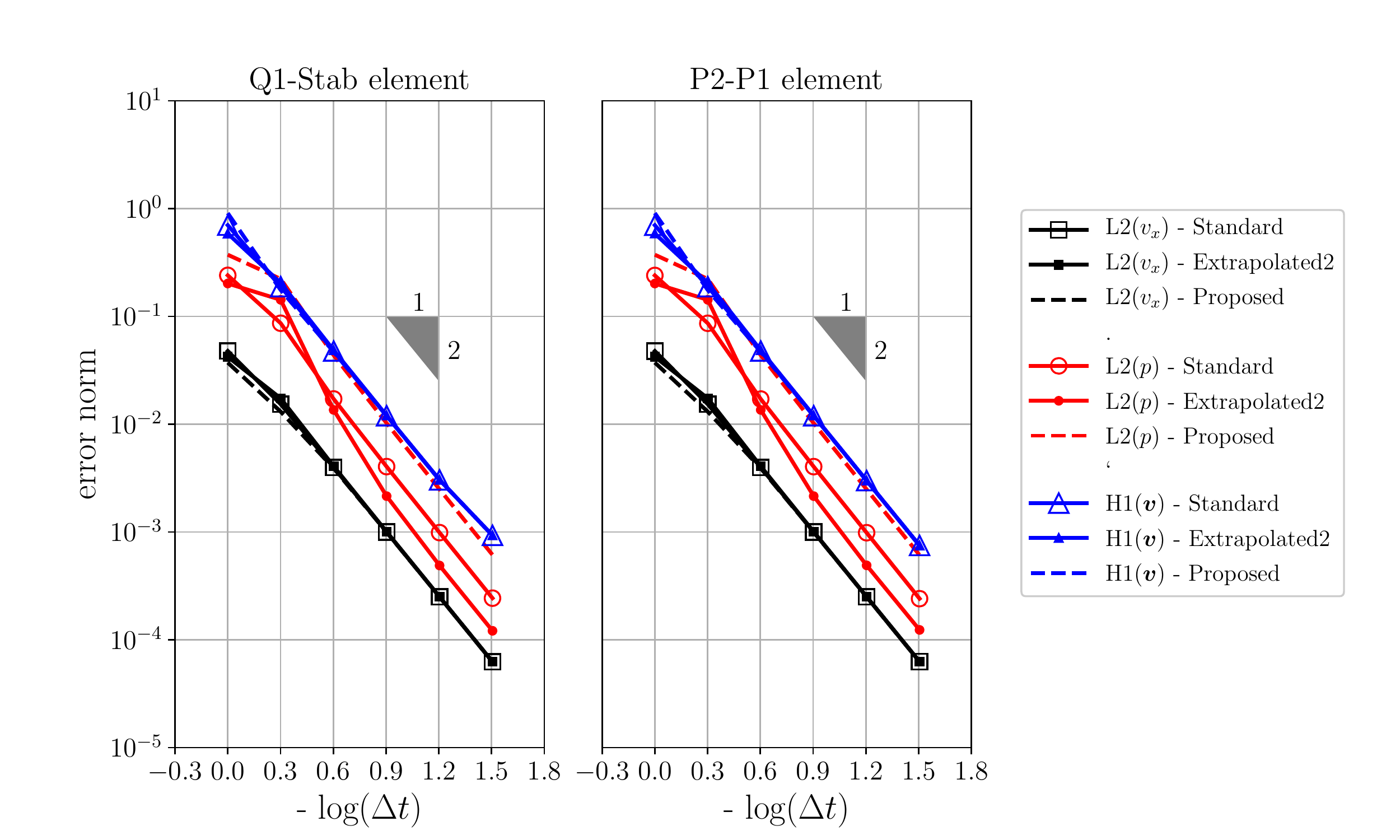}
\caption{Manufactured solution example: convergence of error norms at $t=5$ obtained with GA scheme with $\rho_{\infty}=0.5$ using different formulations.}
\label{manfsoln-graphs-GA2-quad}
\end{figure}

In this example, we consider a test case with a manufactured solution similar to the one proposed by Kim and Moin \cite{KimJCP1985}. The properties of the fluid are: $\rho=1$ and $\mu=0.02$. The velocity and pressure fields are assumed to be
\begin{align}
v_x &= - \, \cos(x) \, \sin(y) \, \sin(2 \, t), \\
v_y &= \sin(x) \, \cos(y) \, \sin(2 \, t), \\
p   &= - \, \frac{1}{4} \, [\cos(2 \, x) + \cos(2 \, y)] \, \sin^2(2 \, t).
\end{align}
The balancing body force vector ($\bm{g}$) is computed by substituting the solution in the momentum equation (\ref{NS-main-eqn1}). The domain is assumed to be a square of side one unit. Spatial discretisations with $250 \times 250$ Q1-Stab elements and $100 \times 100 \times 2$ P2-P1 elements are considered. The error norms of velocity and pressure evaluated at the time instant $t=5$ are displayed in Figs. \ref{manfsoln-graphs-BE-quad}, \ref{manfsoln-graphs-GA1-quad} and \ref{manfsoln-graphs-GA2-quad}, respectively, for the BDF1, GA with $\rho_{\infty}=0$ and GA with $\rho_{\infty}=0.5$. As the BDF2 scheme is spectrally equivalent to the GA scheme with $\rho_{\infty}=0$, the error norms obtained with the BDF2 scheme are identical to those of the GA scheme with $\rho_{\infty}=0$. Therefore, results obtained with the BDF2 scheme are omitted to avoid redundancy.

As shown in Figs. \ref{manfsoln-graphs-BE-quad}, \ref{manfsoln-graphs-GA1-quad} and \ref{manfsoln-graphs-GA2-quad}, numerical results obtained with the linearised convection operators converge with optimal converge rates for both the BDF1 and GA schemes for the Q1-Stab as well as the P2-P1 elements. Except for the largest time step size, there is a negligible difference in the norms of velocity computed using the standard nonlinear scheme and the proposed formulations with linearised convection operators. Although the difference in L2 norm for the pressure obtained with the standard and the proposed scheme is noticeable, the rate of convergence in pressure obtained using the proposed approach is still second-order. This minor loss of accuracy in pressure can be ignored for practical applications, and as illustrated using the unsteady flow examples of flow past a circular cylinder and fluid-structure interaction, the effect of this difference is negligible.

\subsection{Flow past a fixed circular cylinder in 2D}
\begin{figure}[H]
 \begin{center}
 \includegraphics[trim = 0mm 0mm 0mm 0mm, clip, scale=1.4]{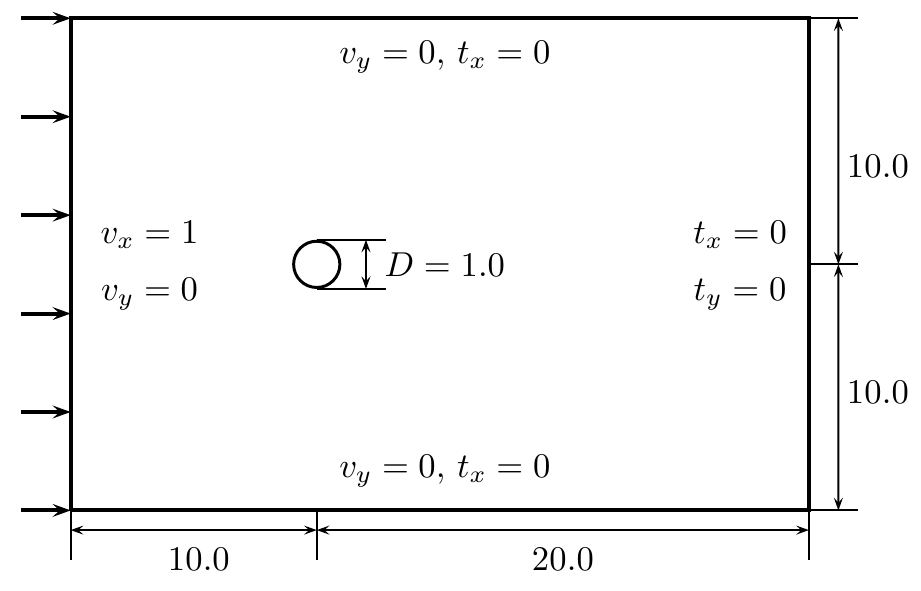}
\caption{Flow past circular cylinder: geometry and boundary conditions. Length units are centimeters.}
\label{fig-cylinder-geom}
 \end{center}
\end{figure}

\begin{figure}[H]
\begin{center}
\includegraphics[trim = 0mm 0mm 0mm 0mm, clip, scale=0.32]{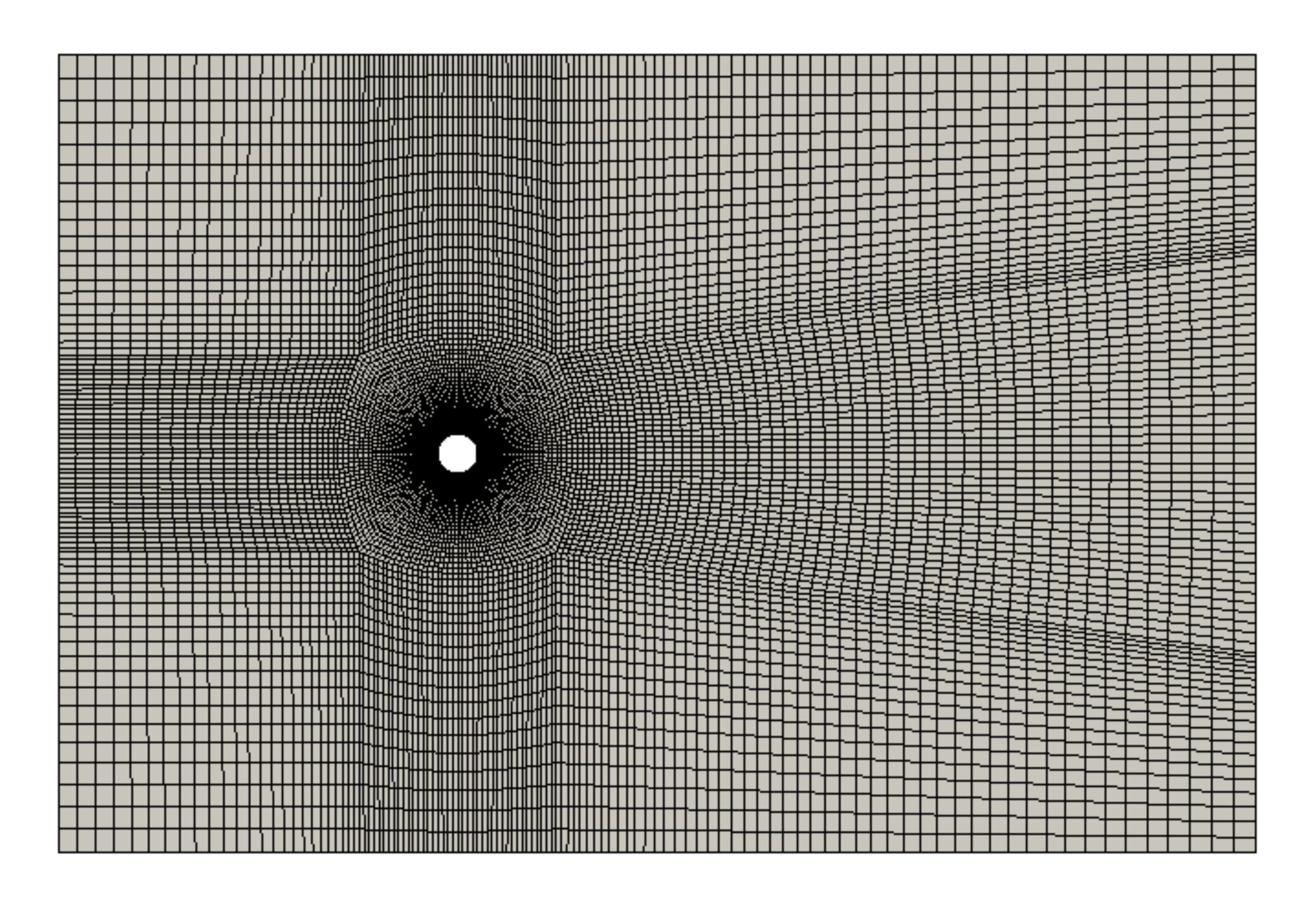}
\includegraphics[trim = 0mm 0mm 0mm 0mm, clip, scale=0.32]{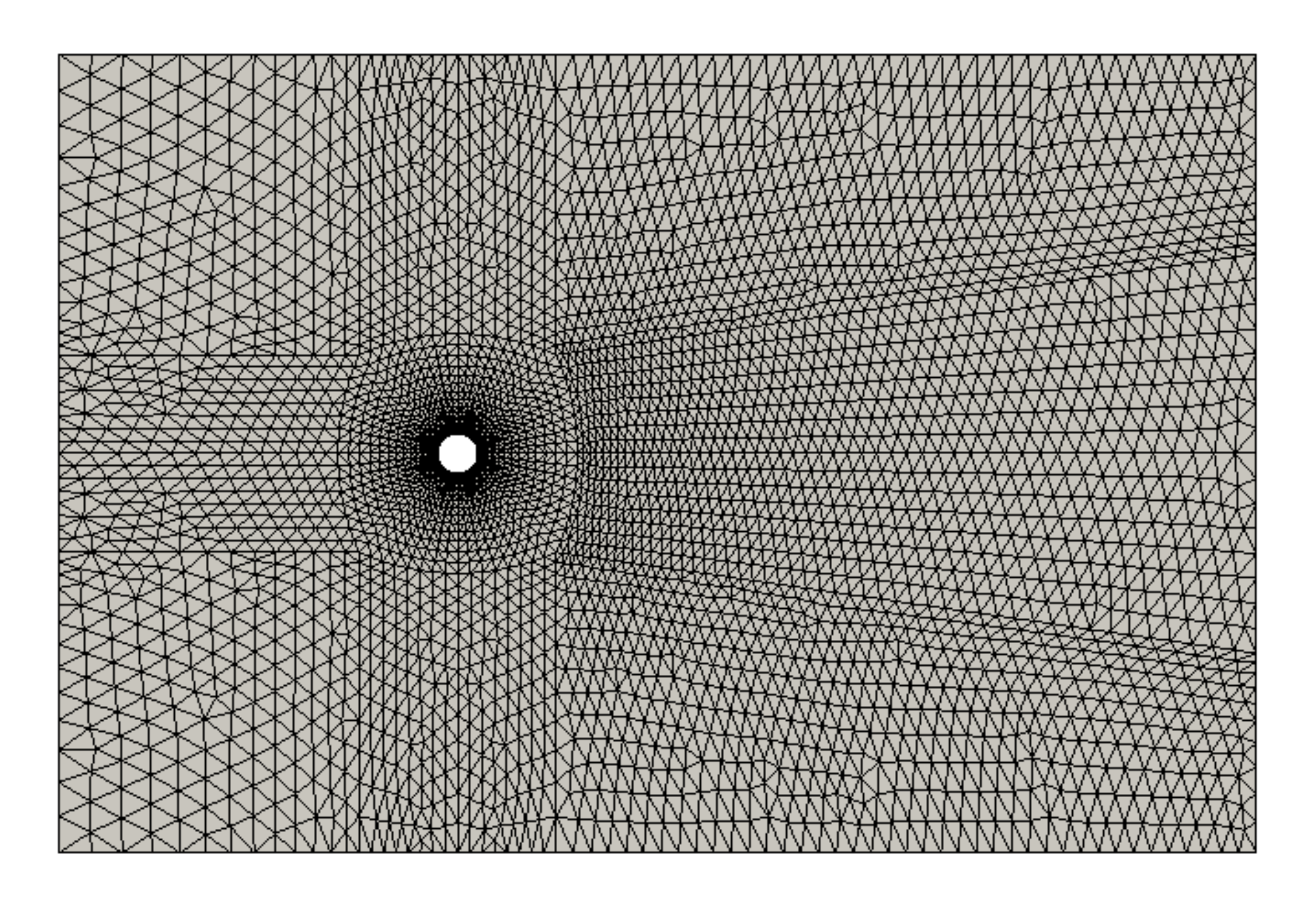}
\caption{Flow past circular cylinder: body-fitted finite element meshes used for the simulations.}
\label{fig-cylinder-meshes}
 \end{center}
\end{figure}

Having established the temporal accuracy of the proposed scheme in the previous example, we now demonstrate the accuracy and the computational advantages of the proposed scheme using the benchmark example of unsteady flow past a fixed circular cylinder. The geometry and boundary conditions of the problem are shown in Fig. \ref{fig-cylinder-geom}. The finite element meshes used for the simulations are shown in Fig. \ref{fig-cylinder-meshes}. The quadrilateral mesh consists of 12675 nodes and 12400 Q1-Stab elements with 160 linear edges along the circumference of the circle. The triangular mesh consists of 15212 velocity nodes, 3873 pressure nodes and 7466 P2-P1 elements with 64 quadratic edges along the circumference of the circle.

The density of the fluid and the inlet velocity are taken as $\rho=1$ g/cm$^3$ and $(v_x,v_y)=(1,0)$ cm/s, respectively. The viscosity of the fluid is adjusted to match the Reynolds number, which is evaluated as $Re=\rho \, v_{\inf} \, D/\mu$, where $D$ is the diameter of the cylinder and $v_{\inf}$ is the free-stream velocity. In the present case, $D=1$ cm and $v_{\inf}=1$ cm/s. In this work, we consider three different Reynolds numbers, 100, 200 and 400, to illustrate the accuracy of the proposed scheme over a wide range of Reynolds numbers. For each Reynolds number, simulations are carried out for four different time steps, $\Delta t=\{ 0.4, 0.2, 0.1, 0.05 \}$ s, with the standard nonlinear and the proposed linear formulations as well as the formulation based on the second-order accurate extrapolated convection velocity. The spectral radius parameter for the generalised-alpha scheme is taken as $\rho_{\infty}=0.0$.

In all the simulations, inlet velocity is ramped linearly from zero to one during the first second. All the simulations are carried out up to 1000 seconds. The accuracy of the results is assessed by computing the amplitude of lift coefficient ($C_{L}$) and Strouhal number ($St$), the vortex shedding frequency.
%

\begin{figure}[H]
 \begin{center}
  \subfigure[$\Delta t = 0.2$] { \includegraphics[trim = 0mm 0mm 0mm 0mm, clip, scale=0.5]{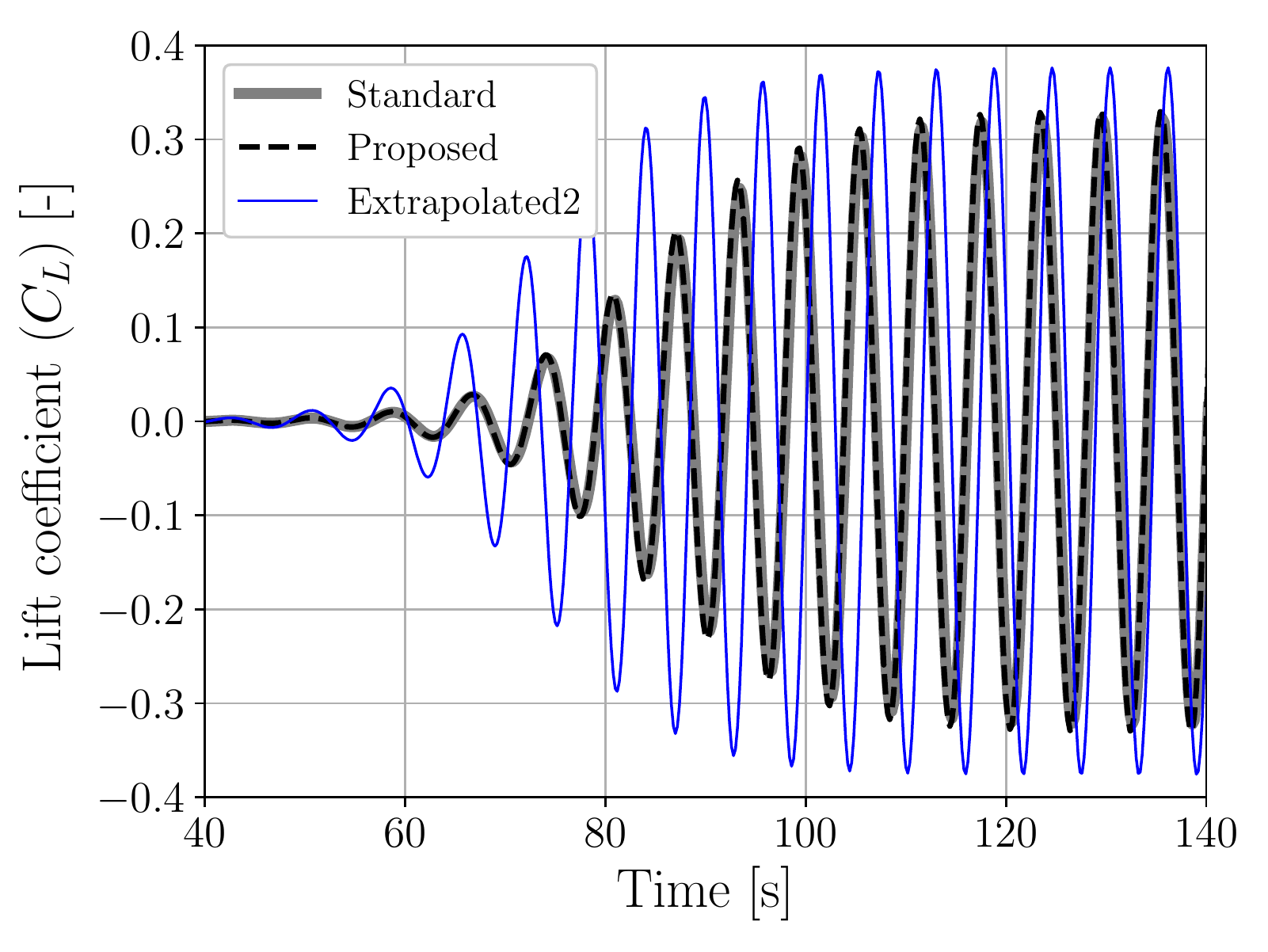}}
  \subfigure[$\Delta t = 0.05$]{ \includegraphics[trim = 0mm 0mm 0mm 0mm, clip, scale=0.5]{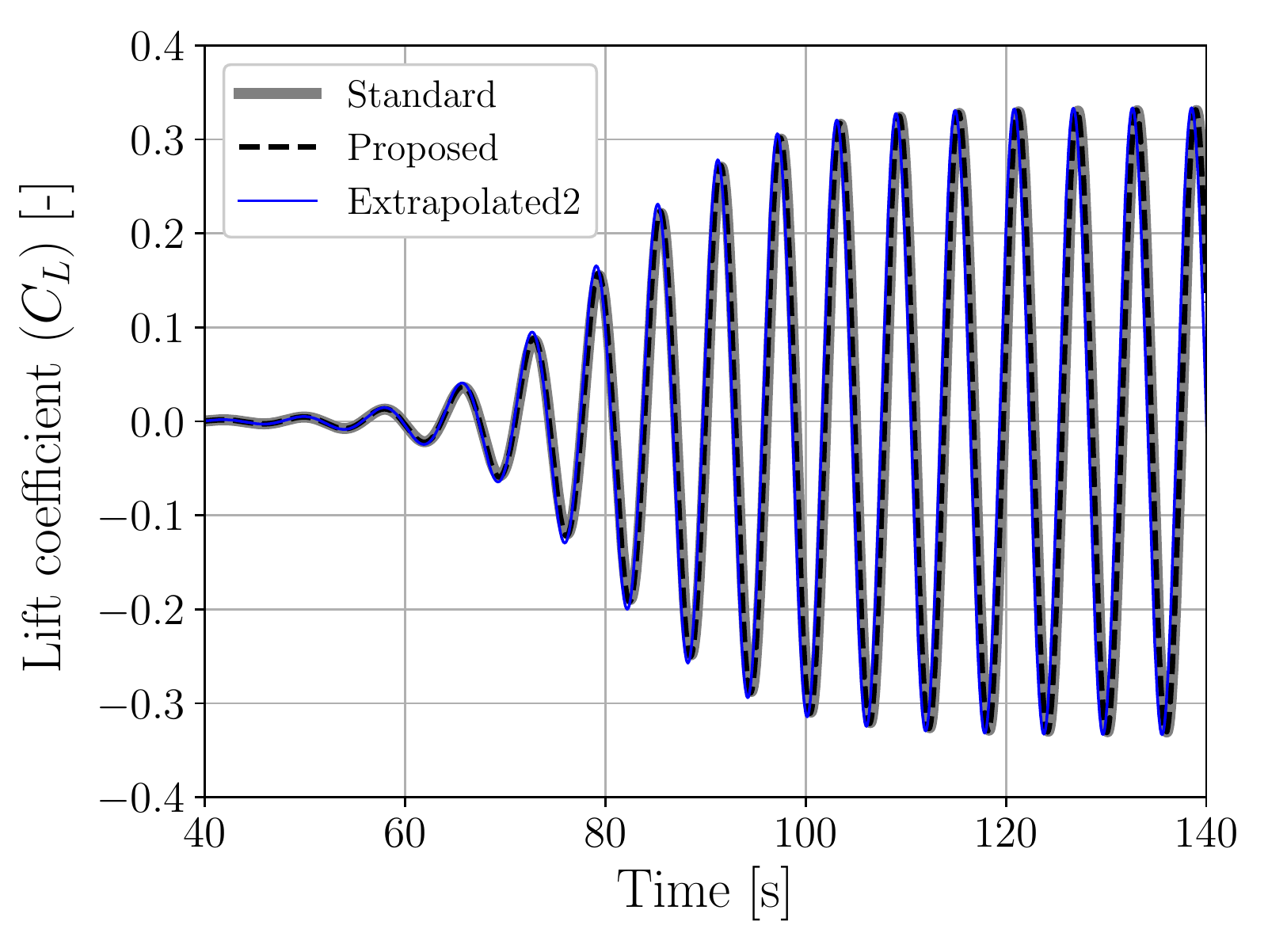}}
 \caption{Flow past circular cylinder: evolution of $C_L$ for $Re=100$ with the Q1-Stab element.}
\label{fig-cylinder-graphs-lift-Re100-quad}
 \end{center}
\end{figure}

\begin{figure}[H]
 \begin{center}
  \subfigure[$\Delta t = 0.2$] { \includegraphics[trim = 0mm 0mm 0mm 0mm, clip, scale=0.5]{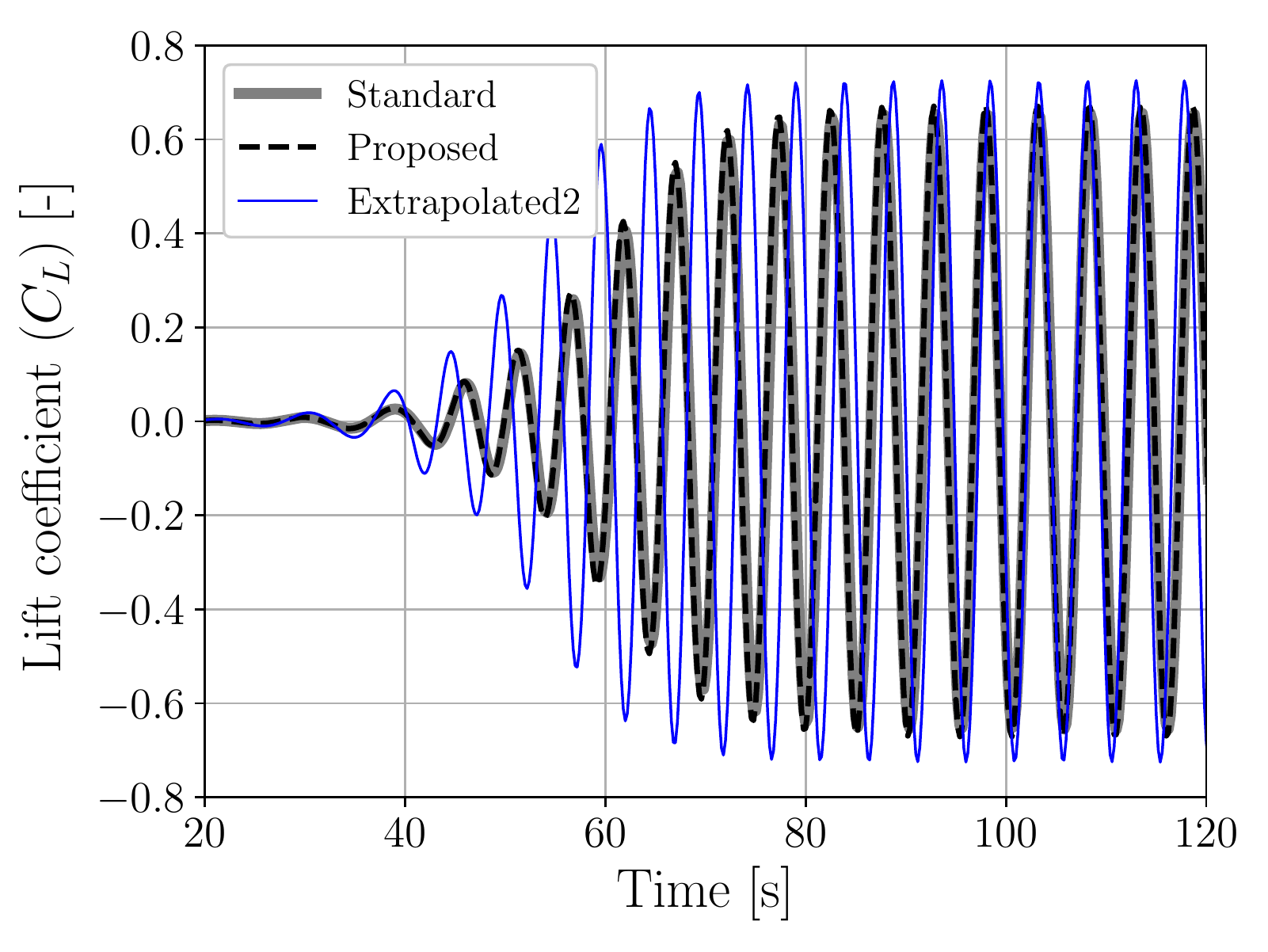}}
  \subfigure[$\Delta t = 0.05$]{ \includegraphics[trim = 0mm 0mm 0mm 0mm, clip, scale=0.5]{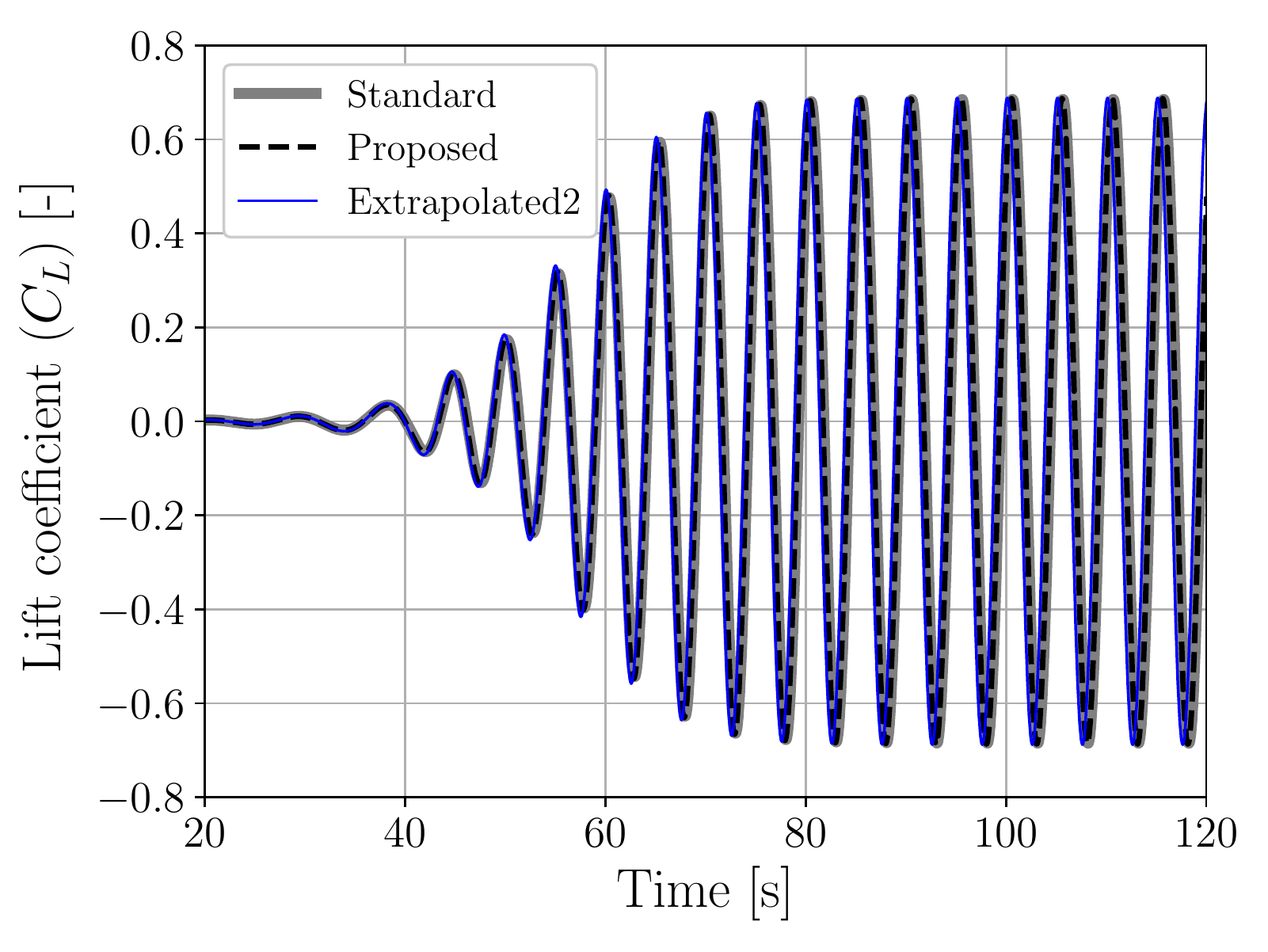}}
 \caption{Flow past circular cylinder: evolution of $C_L$ for $Re=200$ with the Q1-Stab element.}
\label{fig-cylinder-graphs-lift-Re200-quad}
 \end{center}
\end{figure}

\begin{figure}[H]
 \begin{center}
  \subfigure[$\Delta t = 0.2$] { \includegraphics[trim = 0mm 0mm 0mm 0mm, clip, scale=0.5]{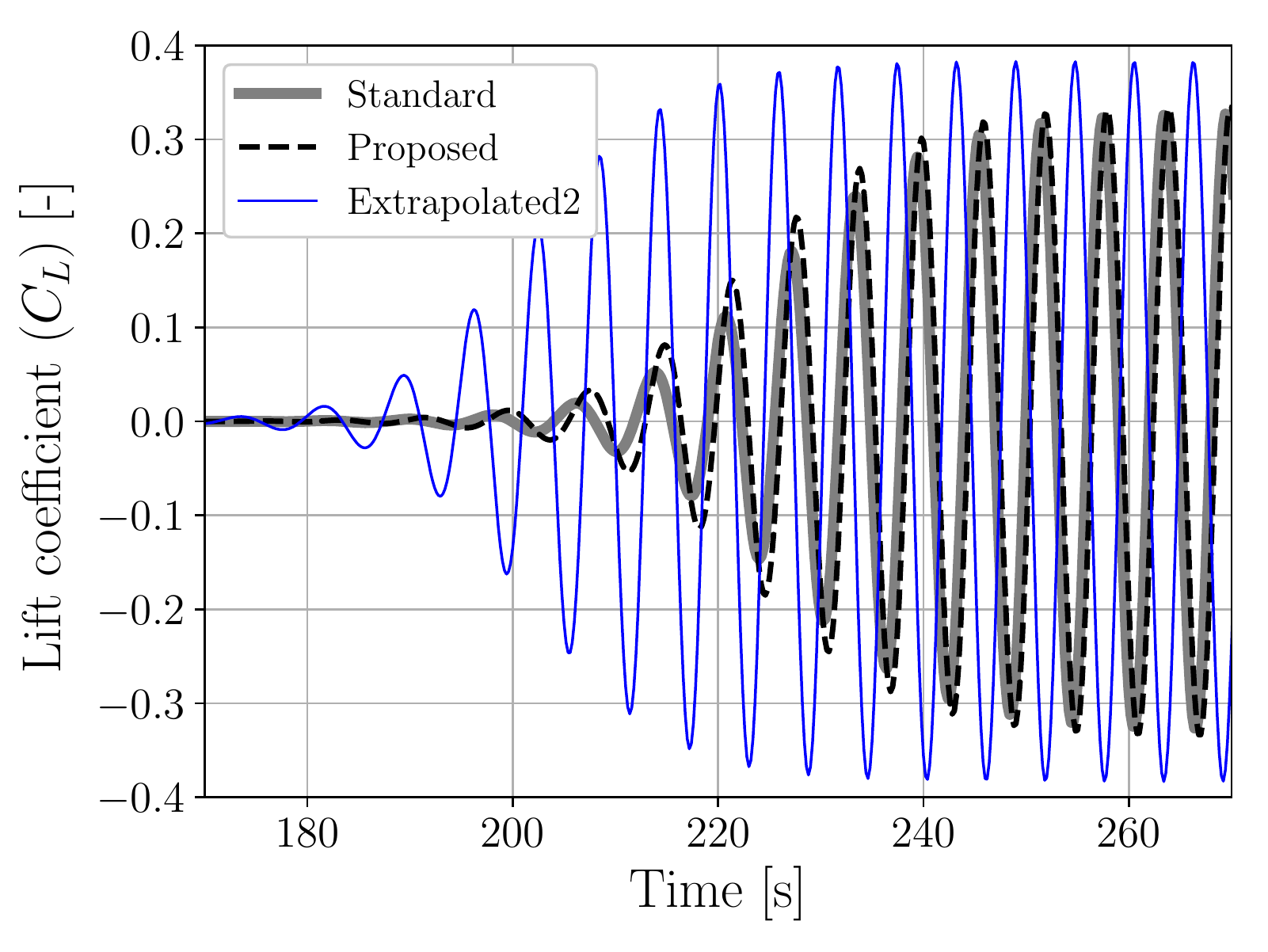}}
  \subfigure[$\Delta t = 0.05$]{ \includegraphics[trim = 0mm 0mm 0mm 0mm, clip, scale=0.5]{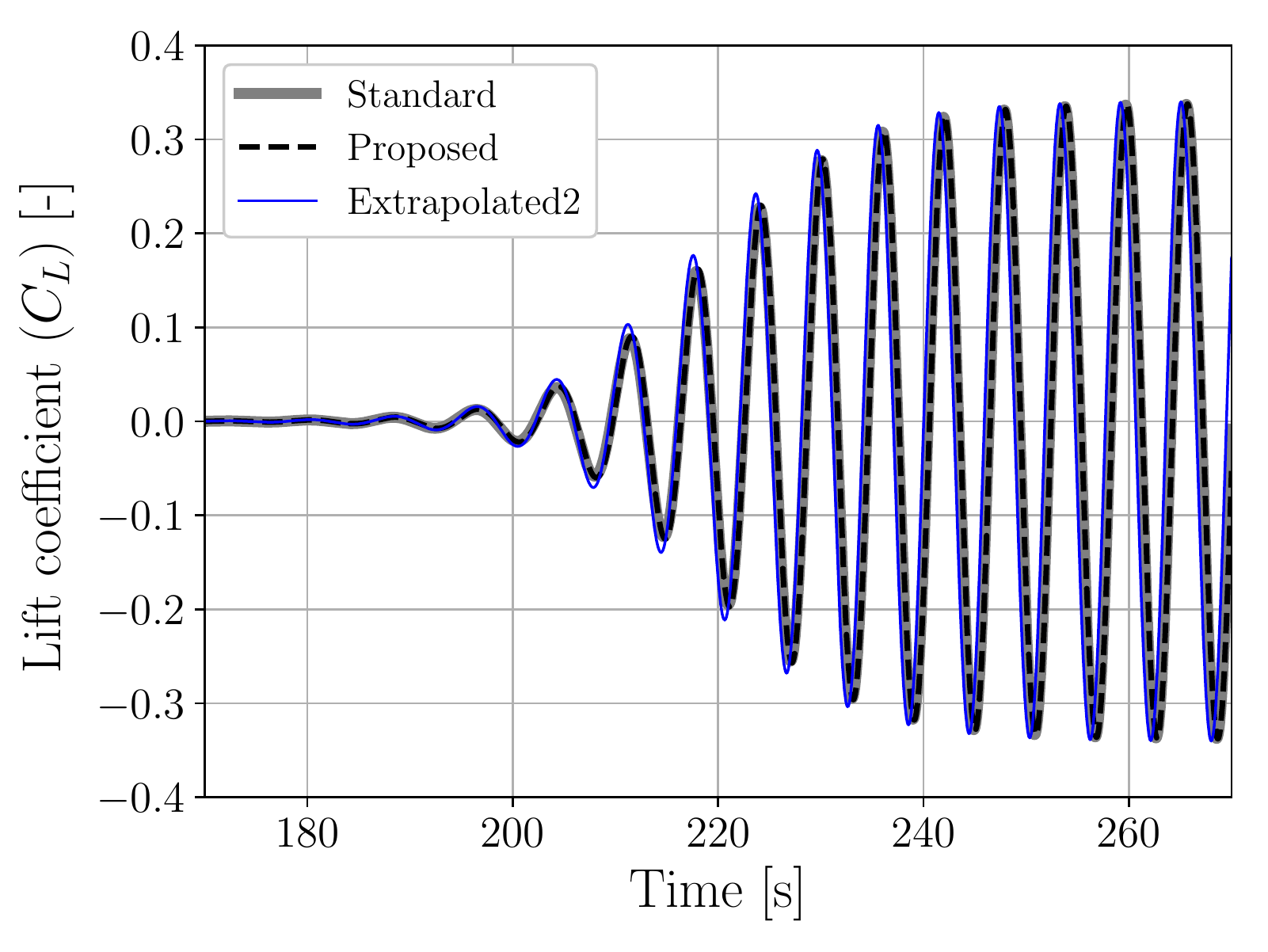}}
 \caption{Flow past circular cylinder: evolution of $C_L$ for $Re=100$ with the P2-P1 element.}
\label{fig-cylinder-graphs-lift-Re100-tria}
 \end{center}
\end{figure}

\begin{figure}[H]
 \begin{center}
  \subfigure[$\Delta t = 0.2$] { \includegraphics[trim = 0mm 0mm 0mm 0mm, clip, scale=0.5]{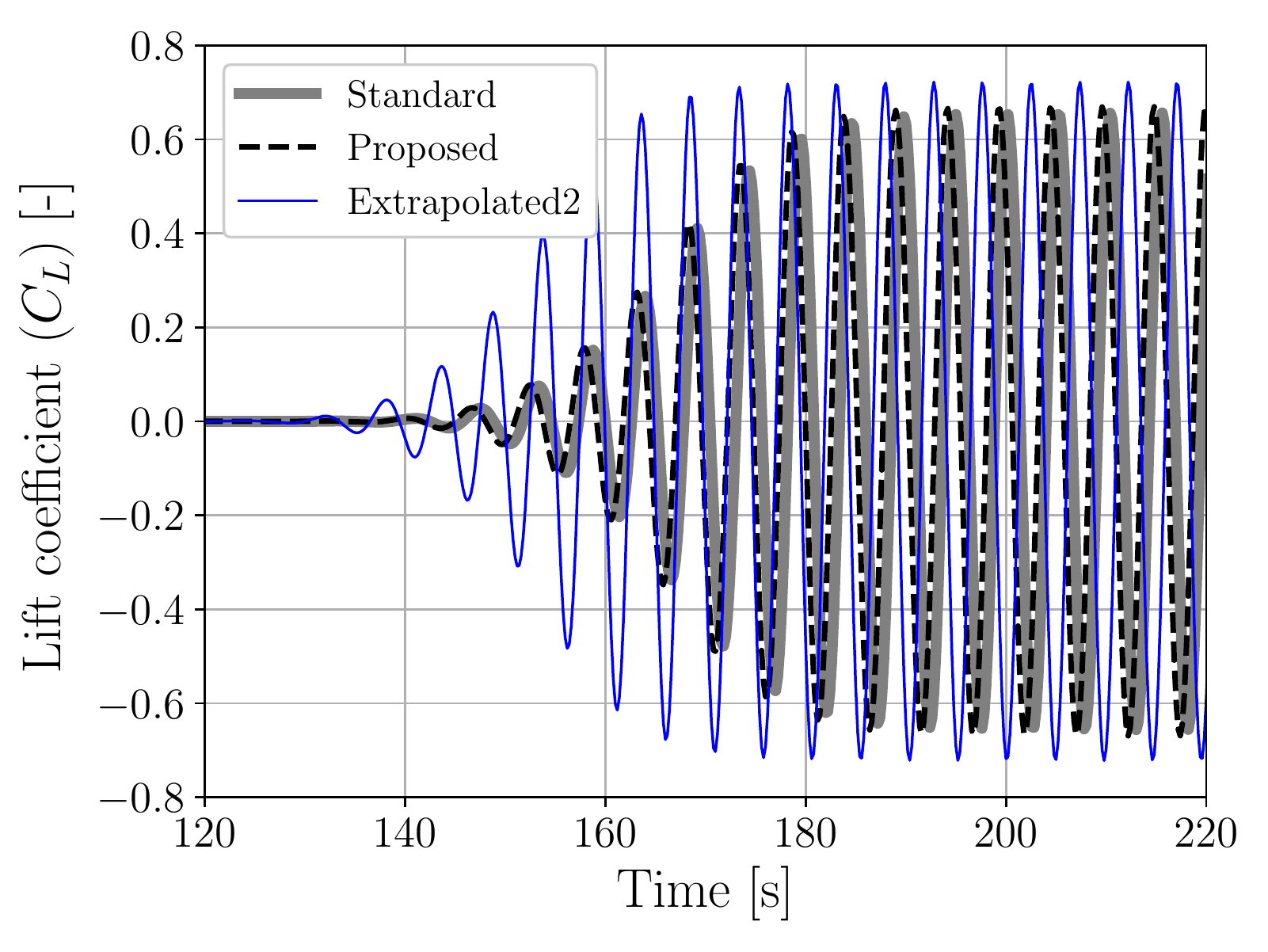}}
  \subfigure[$\Delta t = 0.05$]{ \includegraphics[trim = 0mm 0mm 0mm 0mm, clip, scale=0.5]{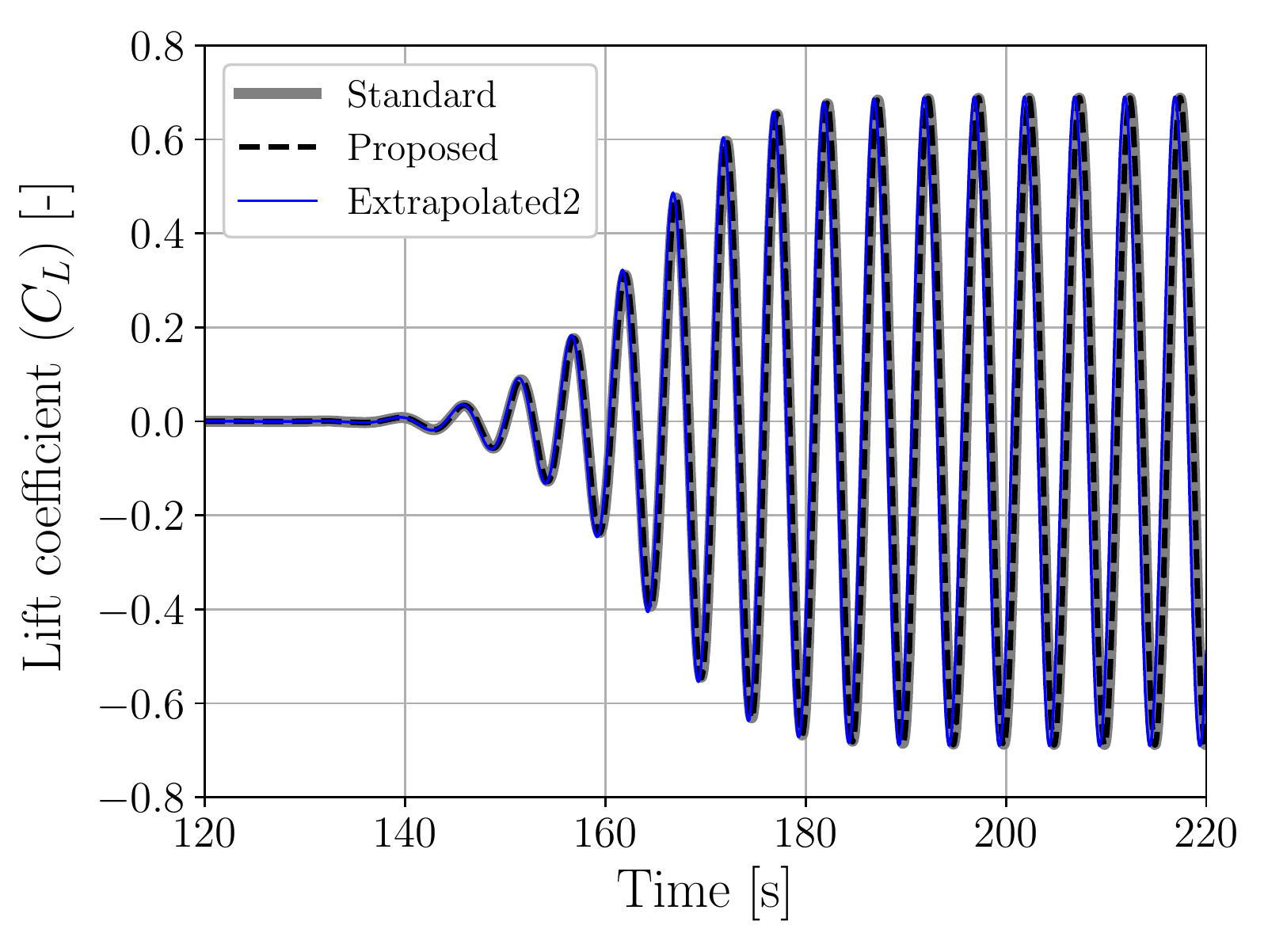}}
 \caption{Flow past circular cylinder: evolution of $C_L$ for $Re=200$ with the P2-P1 element.}
\label{fig-cylinder-graphs-lift-Re200-tria}
 \end{center}
\end{figure}

\begin{figure}[H]
 \begin{center}
  \subfigure[At $t=150$ for $Re=100$ with $\Delta t = 0.1$] { \includegraphics[trim = 0mm 0mm 0mm 0mm, clip, scale=0.45]{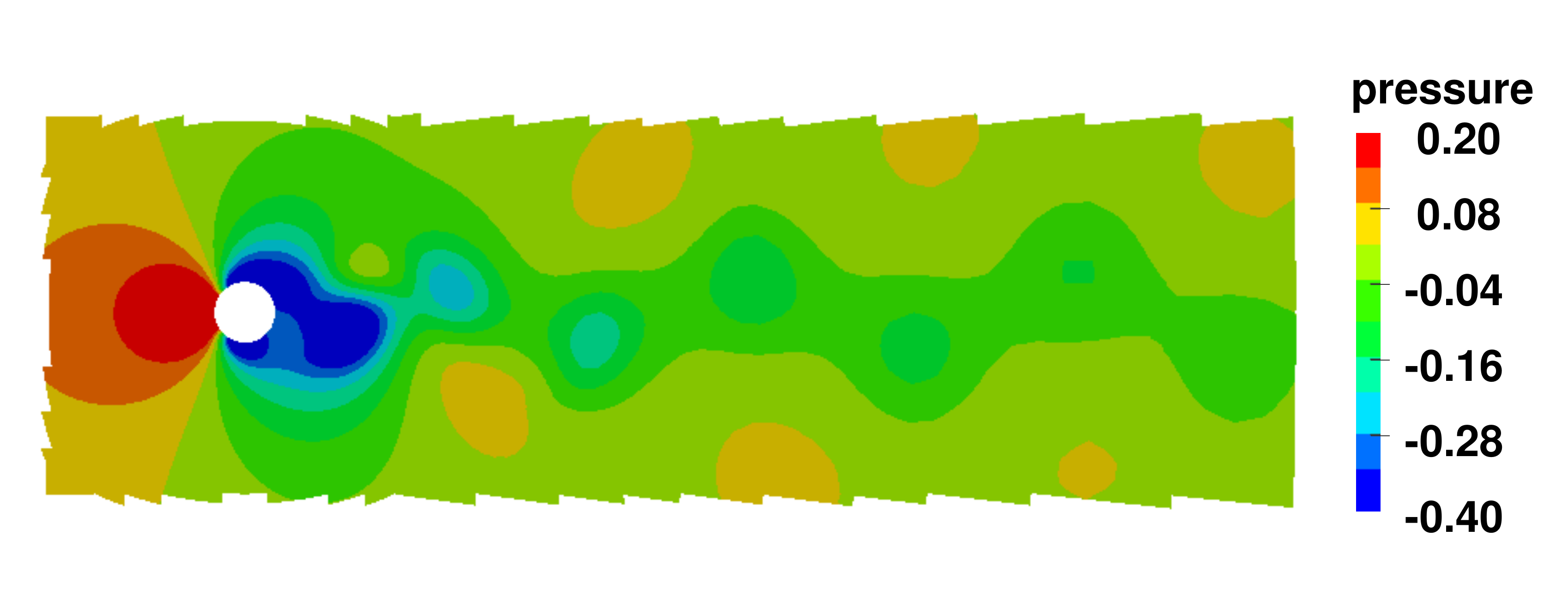}}
  \subfigure[At $t=80$ for $Re=400$ with $\Delta t = 0.1$]{ \includegraphics[trim = 0mm 0mm 0mm 0mm, clip, scale=0.45]{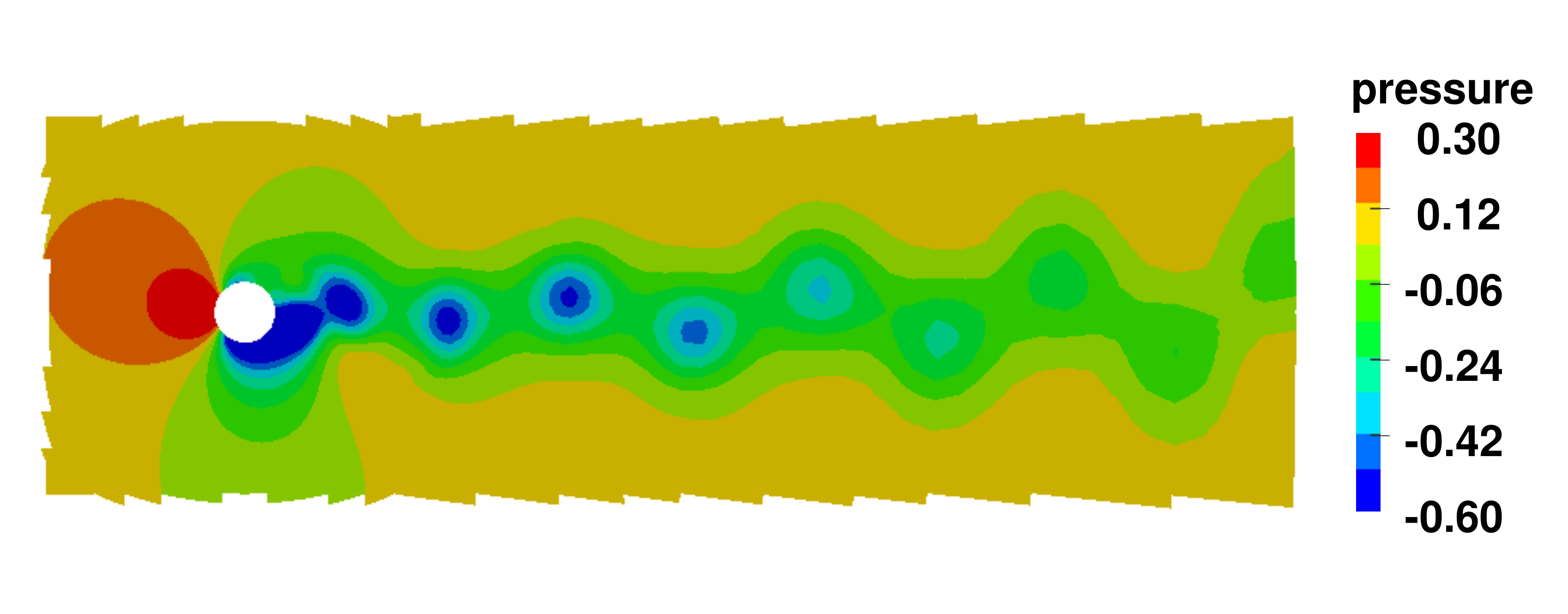}}
 \caption{Flow past circular cylinder: contour plots of pressure obtained with the proposed scheme using the Q1-Stab element.}
\label{fig-cylinder-contours-quad}
 \end{center}
\end{figure}

The computed values of $C_L$ and $St$ are tabulated in Tables. \ref{table-cylinder-Re100}, \ref{table-cylinder-Re200}, \ref{table-cylinder-Re400}, respectively, for Reynolds numbers 100, 200 and 400. The graphs for the evolution of the lift coefficient obtained with the Q1-Stab element with two different time steps are presented in Figs. \ref{fig-cylinder-graphs-lift-Re100-quad} and \ref{fig-cylinder-graphs-lift-Re200-quad}, respectively, for Reynolds numbers 100 and 200. The corresponding graphs for the P2-P1 element are shown in Figs. \ref{fig-cylinder-graphs-lift-Re100-tria} and \ref{fig-cylinder-graphs-lift-Re200-tria}. These results illustrate that numerical results that are in excellent agreement with the standard scheme can be obtained using the proposed iteration-free scheme. While both the proposed and the \emph{Extrapolated2} formulations convergence with $\Delta t$, the agreement with the standard formulation for large time steps is consistently better with the proposed formulation. The contour plots of pressure presented in Fig. \ref{fig-cylinder-contours-quad} show that the pressure field obtained with the proposed scheme is smooth. The slight drop in accuracy in $C_L$ for the P2-P1 element for $Re=400$ is attributed to the coarser spatial discretisation for pressure for the P2-P1 element when compared with the Q1-Stab element.

The computational benefits of the proposed scheme are assessed by studying the number of iterations required for the standard methodology using the iteration-based Newton-Raphson scheme. The average number of iterations required in each simulation using the standard scheme is tabulated in Table. \ref{table-cylinder-iterations}. As expected, the number of required iterations with the standard approach increases with an increase in the value of Reynolds number, especially with large time steps, $\Delta t=0.4$ and $\Delta t=0.2$. On the contrary, the formulations with linearised convection operators require only one iteration at every time step irrespective of the Reynolds number and the time step size. Note that although the proposed and the \emph{Extrapolated2} formulations require only one iteration, the proposed scheme computes numerical results of the same accuracy as that of the standard nonlinear scheme, unlike the \emph{Extrapolated2} formulation which still requires smaller time steps for obtaining results that are comparable to that of the standard nonlinear scheme. Therefore, with its ability to compute numerical results that are in good agreement with the standard scheme even when using large time steps, the proposed scheme yields significant computational benefits over the standard and \emph{Extrapolated2} schemes. It is also worth pointing out that the resulting computational benefits increase for flows with higher Reynolds numbers. In this particular example, 3X-6X speedups are achieved, as can be deduced from Table \ref{table-cylinder-iterations}, with minimal changes to the existing code and without even resorting to the techniques of parallelisation.

\textbf{Remark:} Numerical results with the BDF1 scheme show the same trend as that observed with the GA scheme, i.e., the results obtained with the proposed scheme are in much better agreement with the standard scheme than those computed with the \emph{Extrapolated1} scheme.

\renewcommand{\arraystretch}{1.5}
\begin{table}[H]
\centering
\begin{tabular}{|c|c|c|c|c|c|c|}
\hline
\multirow{2}{0.8cm}{Data} & \multicolumn{3}{|c|}{\multirow{1}{*}{$C_L$}} & \multicolumn{3}{|c|}{\multirow{1}[2]{*}{$St$}} \\
\cline{2-7} &  Standard &  Proposed &  Extrapolated2  &  Standard & Proposed  &  Extrapolated2    \\
\hline
\textcolor{blue}{Present -- Q1-Stab}     &  &  &  &  &  &  \\
$\Delta t = 0.4$    &   0.286   &   0.293   &  0.532   &   0.156   &   0.172   &   0.153   \\
$\Delta t = 0.2$    &   0.333   &   0.331   &  0.376   &   0.168   &   0.174   &   0.168   \\
$\Delta t = 0.1$    &   0.323   &   0.332   &  0.342   &   0.170   &   0.172   &   0.170   \\
$\Delta t = 0.05$   &   0.331   &   0.332   &  0.334   &   0.168   &   0.168   &   0.168   \\
\hline
\textcolor{blue}{Present -- P2-P1}      &  &  &  &  &  &  \\
$\Delta t = 0.4$  &   0.293   &   0.316   &  0.541   &   0.159   &   0.158   &   0.176   \\
$\Delta t = 0.2$  &   0.330   &   0.337   &  0.383   &   0.167   &   0.167   &   0.173   \\
$\Delta t = 0.1$  &   0.337   &   0.339   &  0.349   &   0.169   &   0.169   &   0.167   \\
$\Delta t = 0.05$ &   0.338   &   0.338   &  0.341   &   0.169   &   0.169   &   0.170   \\
\hline
Braza et al. \cite{BrazaJFM1986}            & \multicolumn{3}{c|}{0.300} & \multicolumn{3}{c|}{0.160} \\
Calhoun \cite{CalhounJCP2002}               & \multicolumn{3}{c|}{0.298} & \multicolumn{3}{c|}{0.175} \\
Liu et al. \cite{LiuJCP1998}                & \multicolumn{3}{c|}{0.339} & \multicolumn{3}{c|}{0.165} \\
Kadapa et al. \cite{KadapaCMAME2016fsi}     & \multicolumn{3}{c|}{0.339} & \multicolumn{3}{c|}{0.166} \\
Kadapa et al. \cite{KadapaCMAME2017rigid}   & \multicolumn{3}{c|}{0.330} & \multicolumn{3}{c|}{0.169} \\
\hline
 \end{tabular}
\caption{Flow past circular cylinder: $C_L$ and $St$ values for $Re=100$.}
\label{table-cylinder-Re100}
\end{table}
\begin{table}[H]
\centering
\begin{tabular}{|c|c|c|c|c|c|c|}
\hline
\multirow{2}{0.8cm}{Data} & \multicolumn{3}{|c|}{\multirow{1}{*}{$C_L$}} & \multicolumn{3}{|c|}{\multirow{1}[2]{*}{$St$}} \\
\cline{2-7} &  Standard  &  Proposed &  Extrapolated2  &  Standard  &  Proposed &  Extrapolated2    \\
\hline
\textcolor{blue}{Present -- Q1-Stab}     &  &  &  &  &  &  \\
$\Delta t = 0.4$  &   0.546   &   0.579   &  0.892  &  0.172   &   0.180  &  0.198 \\
$\Delta t = 0.2$  &   0.656   &   0.672   &  0.726  &  0.194   &   0.194  &  0.206 \\
$\Delta t = 0.1$  &   0.682   &   0.688   &  0.696  &  0.198   &   0.198  &  0.202 \\
$\Delta t = 0.05$ &   0.685   &   0.687   &  0.688  &  0.198   &   0.198  &  0.200 \\
\hline
\textcolor{blue}{Present -- P2-P1}      &  &  &  &  &  &  \\
$\Delta t = 0.4$   &   0.544   &   0.573   &  0.885   &   0.181   &   0.180  &  0.199 \\
$\Delta t = 0.2$   &   0.656   &   0.670   &  0.723   &   0.194   &   0.193  &  0.204 \\
$\Delta t = 0.1$   &   0.689   &   0.688   &  0.698   &   0.202   &   0.197  &  0.202 \\
$\Delta t = 0.05$  &   0.688   &   0.689   &  0.691   &   0.202   &   0.202  &  0.199 \\
\hline
Braza et al. \cite{BrazaJFM1986}           & \multicolumn{3}{c|}{0.750} & \multicolumn{3}{c|}{0.200} \\
Franke et al. \cite{FrankeJWEIA1990}       & \multicolumn{3}{c|}{0.650} & \multicolumn{3}{c|}{0.194} \\
Calhoun \cite{CalhounJCP2002}              & \multicolumn{3}{c|}{0.668} & \multicolumn{3}{c|}{0.202} \\
Liu et al. \cite{LiuJCP1998}               & \multicolumn{3}{c|}{0.690} & \multicolumn{3}{c|}{0.192} \\
Kadapa et al. \cite{KadapaCMAME2016fsi}    & \multicolumn{3}{c|}{0.711} & \multicolumn{3}{c|}{0.194} \\
Kadapa et al. \cite{KadapaCMAME2017rigid}  & \multicolumn{3}{c|}{0.717} & \multicolumn{3}{c|}{0.196} \\
\hline
 \end{tabular}
\caption{Flow past circular cylinder: $C_L$ and $St$ values for $Re=200$.}
\label{table-cylinder-Re200}
\end{table}
\begin{table}[H]
\centering
\begin{tabular}{|c|c|c|c|c|c|c|}
\hline
\multirow{2}{0.8cm}{Data} & \multicolumn{3}{|c|}{\multirow{1}{*}{$C_L$}} & \multicolumn{3}{|c|}{\multirow{1}[2]{*}{$St$}} \\
\cline{2-7} &  Standard  &  Proposed &  Extrapolated2  &  Standard  &  Proposed &  Extrapolated2    \\
\hline
\textcolor{blue}{Present -- Q1-Stab}    &  &  &  &  &  &  \\
$\Delta t = 0.4$   &   0.827   &  0.848    &   1.167  &   0.196   &   0.190   &   0.206   \\
$\Delta t = 0.2$   &   1.006   &  1.023    &   1.063  &   0.215   &   0.215   &   0.232   \\
$\Delta t = 0.1$   &   1.052   &  1.061    &   1.064  &   0.218   &   0.218   &   0.224   \\
$\Delta t = 0.05$  &   1.063   &  1.066    &   1.066  &   0.220   &   0.220   &   0.222   \\
\hline
\textcolor{blue}{Present -- P2-P1}     &  &  &  &  &  &  \\
$\Delta t = 0.4$   &   0.707   &  0.717    &   1.013  &   0.197   &   0.196   &   0.221   \\
$\Delta t = 0.2$   &   0.880   &  0.895    &   0.897  &   0.216   &   0.214   &   0.235   \\
$\Delta t = 0.1$   &   0.922   &  0.929    &   0.928  &   0.221   &   0.221   &   0.227   \\
$\Delta t = 0.05$  &   0.930   &  0.932    &   0.931  &   0.223   &   0.223   &   0.224   \\
\hline
Rajani et al. \cite{RajaniAMM2009}      & \multicolumn{3}{c|}{1.000} & \multicolumn{3}{c|}{0.2348} \\
\hline
\end{tabular}
\caption{Flow past circular cylinder: $C_L$ and $St$ values for $Re=400$.}
\label{table-cylinder-Re400}
\end{table}
\renewcommand{\arraystretch}{1.0}
%
%
%
\renewcommand{\arraystretch}{1.5}
\begin{table}[H]
\centering
\begin{tabular}{|c|c|c|c|c|c|c|}
\hline
\multirow{2}{3cm}{Time step ($\Delta t$)} & \multicolumn{3}{|c|}{\multirow{1}{*}{Q1-Stab element}} & \multicolumn{3}{|c|}{\multirow{1}[2]{*}{P2-P1 element}} \\
\cline{2-7} &  $Re=100$  &  $Re=200$ &  $Re=400$  &  $Re=100$  &  $Re=200$ &  $Re=400$    \\
\hline
0.4   &   4.0   &   5.0   &   5.7   &   3.7   &   4.7   &   4.8   \\
0.2   &   4.0   &   4.6   &   5.2   &   3.7   &   3.9   &   4.3   \\
0.1   &   4.0   &   4.0   &   4.9   &   3.7   &   3.9   &   3.9   \\
0.05  &   3.9   &   4.0   &   4.3   &   2.9   &   3.7   &   3.8   \\
\hline
\end{tabular}
\caption{Flow past circular cylinder: average number of iterations for the standard scheme for 1000 seconds in each simulation.}
\label{table-cylinder-iterations}
\end{table}
\renewcommand{\arraystretch}{1.0}

\section{Numerical examples - fluid-structure interaction}  \label{section-examples-fsi}
In this section, we illustrate the computational advantages of the proposed scheme in the context of fluid-structure interaction problems by studying two benchmark examples: (i) a thin flexible restrictor flap in a converging channel \cite{NeumannCSHPC2006} and (ii) vortex-induced vibrations of a thick flexible beam \cite{TurekFSIflex2006}. The simulation platform used in present work is the hierarchical b-spline-based immersed boundary framework established in Dettmer et al. \cite{DettmerCMAME2016} and Kadapa et al. \cite{KadapaCMAME2018, KadapaCMAME2017rigid}. We discuss only the important aspects of the FSI framework in the present work and refer the reader to Kadapa \cite{KadapaCMAME2018} for the comprehensive details.

In the adapted FSI framework, the fluid-solid coupling is resolved using the \emph{iteration-free} staggered scheme proposed by Dettmer and Peri\'c \cite{DettmerIJNME2013}; the fluid problem is discretised with hierarchical b-splines and solved using the mixed velocity-pressure formulation with the SUPG/PSPG stabilisation, and the solid is modelled with first-order quadrilateral/hexahedral elements with the standard displacement or the $\bm{F}$-bar formulation of de Souza Neto et al. \cite{NetoIJSS1996}. The time integration scheme used for the fluid solver is the same generalised-alpha scheme \cite{JansenCMAME2000} that is considered in this paper and the time integration scheme for the elastodynamics is the generalised-alpha scheme proposed by Chung and Hulbert \cite{CHalphaJAM1993}. Both the time integration schemes, as well as the staggered scheme, are \emph{second-order accurate}. The spectral radius parameters for the generalised-alpha schemes for both the solid and fluid problems considered in the following examples are $\rho^s_{\infty}=\rho^f_{\infty}=0$.

Since the proposed formulation has been found to yield better results than the \emph{Extrapolated2} formulation when compared with the standard scheme for the example of flow of a fixed circular cylinder and, more importantly, since the \emph{Extrapolated2} formulation requires the velocity field at $t_{n-1}$ which is not readily available for the cut-cell-based formulation used in this work, only the standard and the proposed formulations are considered for the FSI examples.

The important steps involved in the fluid-structure interaction algorithm adapted in the present work are outlined in Algorithm \ref{algorithm-fsi}, in which $\mathbf{F}_{n-1}$, $\mathbf{F}_{n}$ and $\mathbf{F}_{n+1}$ are the global force vectors on the solid domain at time instants $t_{n-1}$, $t_{n}$ and $t_{n+1}$, respectively;  $\mathbf{F}^{s^P}_{n+1}$ is the predicted force vector on the solid at $t_{n+1}$; $\mathbf{F}^{f}_{n+1}$ is the force exerted by the fluid at $t_{n+1}$; and $\beta$ is the relaxation parameter.

The important detail that is worth pointing out at this point is the computational efficiency of the overall FSI algorithm adapted in the present work. In addition to the iteration-free staggered scheme of \cite{DettmerIJNME2013} which yields significant computational benefits over iteration-based coupling algorithms c.f. \cite{NeumannCSHPC2006, DegrooteCS2009}, the proposed approach increases the computational efficiency further by reducing the computational cost incurred in the fluid solver, Step 4 in Algorithm \ref{algorithm-fsi}. The only iterative scheme used in the entire FSI framework adapted in the present work is for the solid domain for which the computational cost in the majority of FSI problems is only a fraction of the total computational cost of the simulation. Since the fluid solver is the most time-consuming part in the majority of FSI simulations, the overall computational benefits in using the iteration-free coupling scheme and the iteration-free fluid solver are substantial.
\renewcommand{\baselinestretch}{1.5}
\begin{algorithm}[h]
\caption{Algorithm for FSI problem}\label{algorithm-fsi}
\begin{algorithmic}
\State \textbf{Set}: parameters and initial conditions
\State \textbf{Time loop}
   \State \hspace{10mm} \textbf{Step 1}: predict force on the solid: $\mathbf{F}^{s^P}_{n+1} = 2 \, \mathbf{F}_n - \mathbf{F}_{n-1}$
   \State \hspace{10mm} \textbf{Step 2}: solve the solid problem using $\mathbf{F}^{s^P}_{n+1}$
   \State \hspace{10mm} \textbf{Step 3}: reposition immersed solid(s) and update the fluid mesh
   \State \hspace{10mm} \textbf{Step 4}: solve the fluid problem (\emph{\textcolor{blue}{no iterations here}})
   \State \hspace{10mm} \textbf{Step 5}: obtain the fluid force $\mathbf{F}^f_{n+1}$
   \State \hspace{10mm} \textbf{Step 6}: average the interface force: $\mathbf{F}_{n+1} = - \beta \, \mathbf{F}^f_{n+1} + (1-\beta) \, \mathbf{F}^{s^P}_{n+1}$
   \State \hspace{10mm} \textbf{Step 7}: copy variables at $t_{n+1}$ into variables at $t_n$
\State \textbf{End Time loop}
\end{algorithmic}
\end{algorithm}
\renewcommand{\baselinestretch}{1.0}

\subsection{Thin flexible restrictor flap in a converging channel}
\begin{figure}[H]
\centering
\includegraphics[trim = 0mm 0mm 0mm 0mm, clip, scale=0.9]{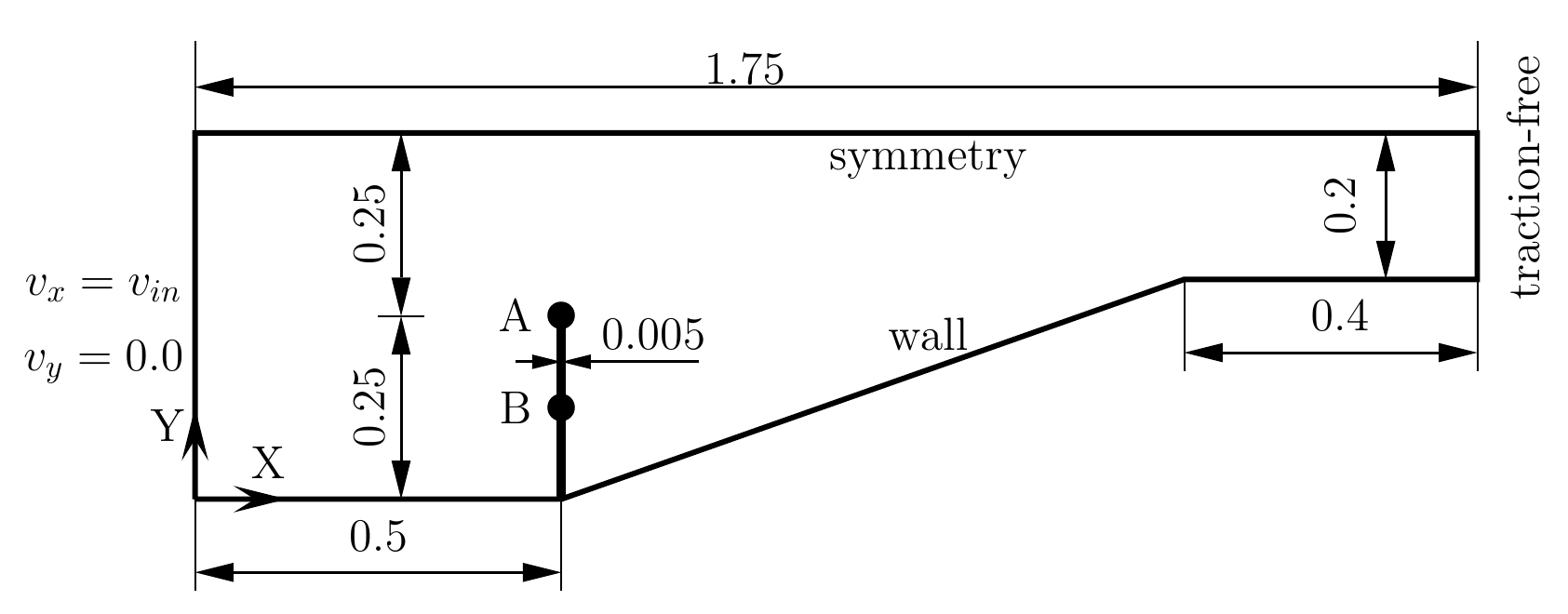}
\caption{Flexible restrictor flap: geometry and boundary conditions. The dimensions are in meters.}
\label{fig-flap-geom}
\end{figure}

This example consists of a thin flexible restrictor flap fixed to the bottom of a converging channel right before the contraction, as depicted schematically in Fig. \ref{fig-flap-geom}. This example was first proposed by Neumann et al. \cite{NeumannCSHPC2006} and was later studied by Degroote et al. \cite{DegrooteCS2009} for evaluating the performance of staggered or partitioned approached to fluid-structure interaction problems with strong added-mass.

Following \cite{DegrooteCS2009}, the Young's modulus and and Poisson's ratio of the flap are taken, respectively, as $E^s = 2.3 \times 10^6$ N/m$^2$ and $\nu^s = 0.45$; the density of the flap's material is taken as $\rho^s=1500$ kg/m$^3$. The material model is assumed to be the Saint Venant-Kirchhoff model. The flap is modelled using $4 \times 100$ bilinear quadrilateral elements with the $\bm{F}$-bar formulation of de Souza Neto et al. \cite{NetoIJSS1996}. The dynamic viscosity of the fluid is $\mu=0.1$ Pa s and the density of the fluid is taken $\rho=\rho^f=1750$ kg/m$^3$. With $\rho^s/\rho^f \approx 0.8571$, this test case corresponds to the extreme case of added-mass in Degroote et al. \cite{DegrooteCS2009}.

\begin{figure}[H]
\centering
\includegraphics[trim = 0mm 0mm 0mm 0mm, clip, scale=0.3]{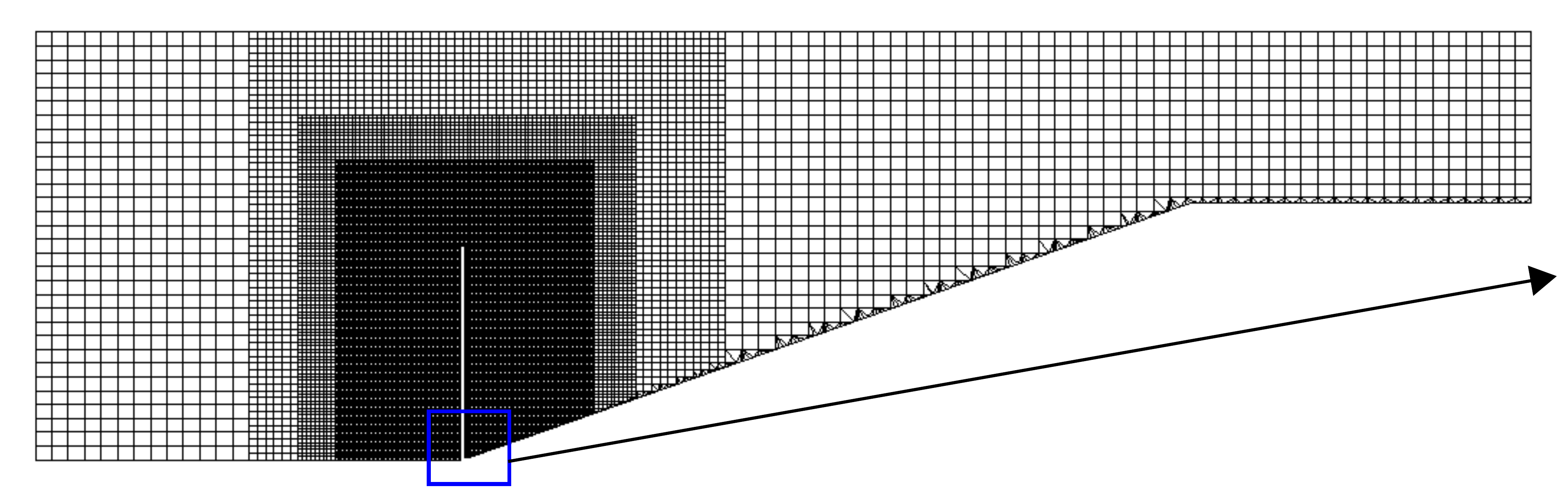}
\includegraphics[trim = 0mm 0mm 0mm 0mm, clip, scale=0.15]{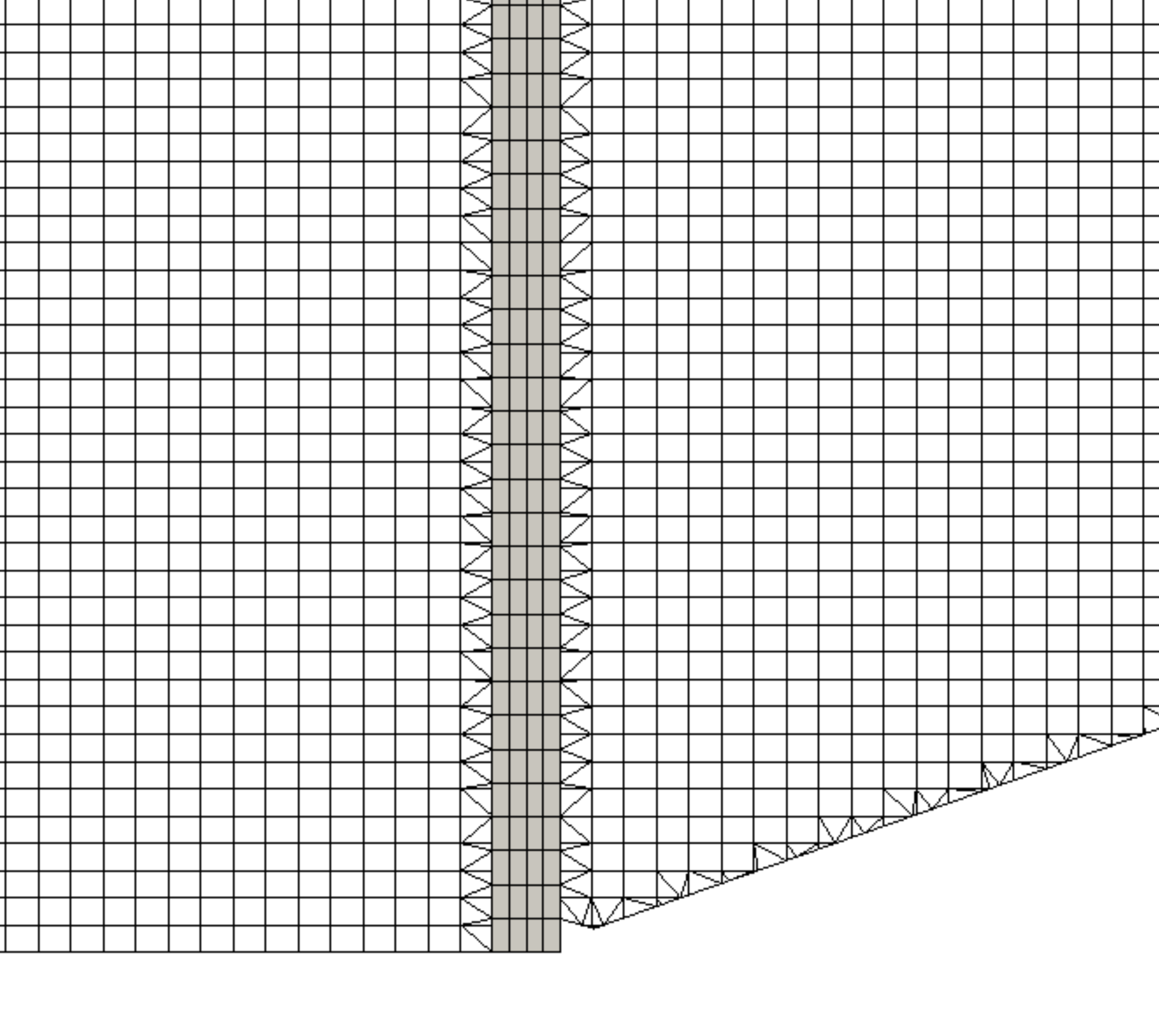}
\caption{Flexible restrictor flap: the fluid mesh at the start of the simulation. Nodes corresponding to points A and B are taken at the middle section of the solid beam model.}
\label{fig-flap-mesh}
\end{figure}

The fluid mesh at the start of the simulation is shown in Fig. \ref{fig-flap-mesh}. As shown, the fluid domain in the vicinity of the flap is refined locally to accommodate the thin flexible flap. With equal-order linear b-splines for velocity and pressure, the number of fluid degrees of freedom at the first time step is 78630, and it remains approximately the same during the whole simulation. Note that it is possible to model this problem using fewer DOFs using an FSI framework that supports beam models for thin solids, for example, Kadapa et al. \cite{KadapaCMAME2016fsi}, Degroote et al. \cite{DegrooteCS2009} and Muddle et al. \cite{MuddleJCP2012}.

The horizontal velocity profile at the inlet ($v_{in}$) is given as
\renewcommand{\arraystretch}{1.5}
\begin{equation}
 v_{in} = \left \{
 \begin{array}{ll}
      0.5 \, V_{max} \, [1 - \cos(\pi \, t/10)], & \textrm{if} \quad 0 \leqslant t \leqslant 10 \\
      V_{max}, & \textrm{if} \quad t \geqslant 10.
    \end{array}
  \right.
\end{equation}
\renewcommand{\arraystretch}{1.0}
where $V_{max}=0.06067$ m/s. The relaxation parameter for the staggered scheme is $\beta=0.01$.

\begin{figure}[H]
\centering
\includegraphics[trim = 0mm 0mm 0mm 0mm, clip, scale=0.6]{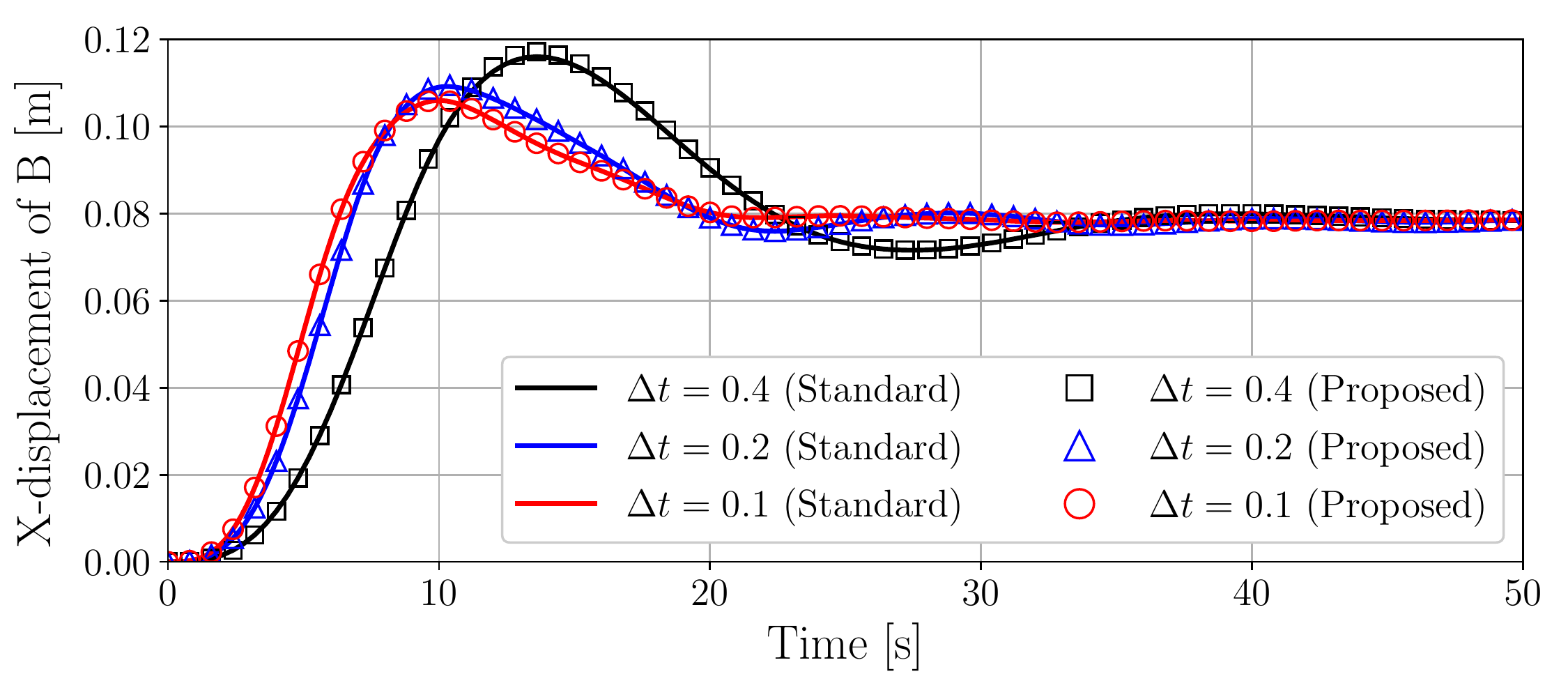}
\includegraphics[trim = 0mm 0mm 0mm 0mm, clip, scale=0.6]{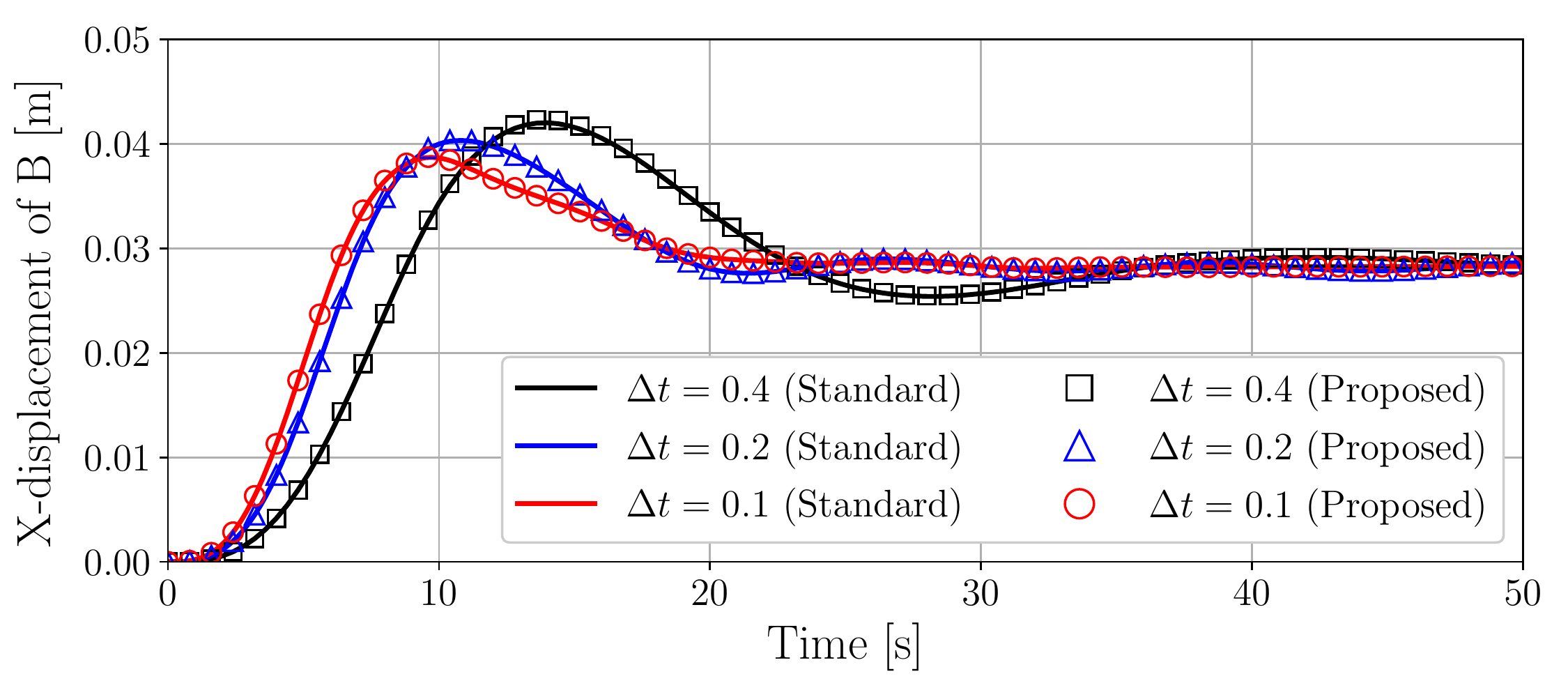}
\caption{Flexible restrictor flap: evolution of X-displacement of points A and B. Note that, for the results obtained with the proposed scheme, only every second, fourth and eighth values are marked, respectively, for $\Delta t=0.4$, $\Delta t=0.2$ and $\Delta t=0.1$.}
\label{fig-flap-graphs-disp}
\end{figure}

\begin{figure}[H]
\centering
\subfigure[Standard scheme]{\includegraphics[trim = 0mm 0mm 0mm 0mm, clip, scale=0.275]{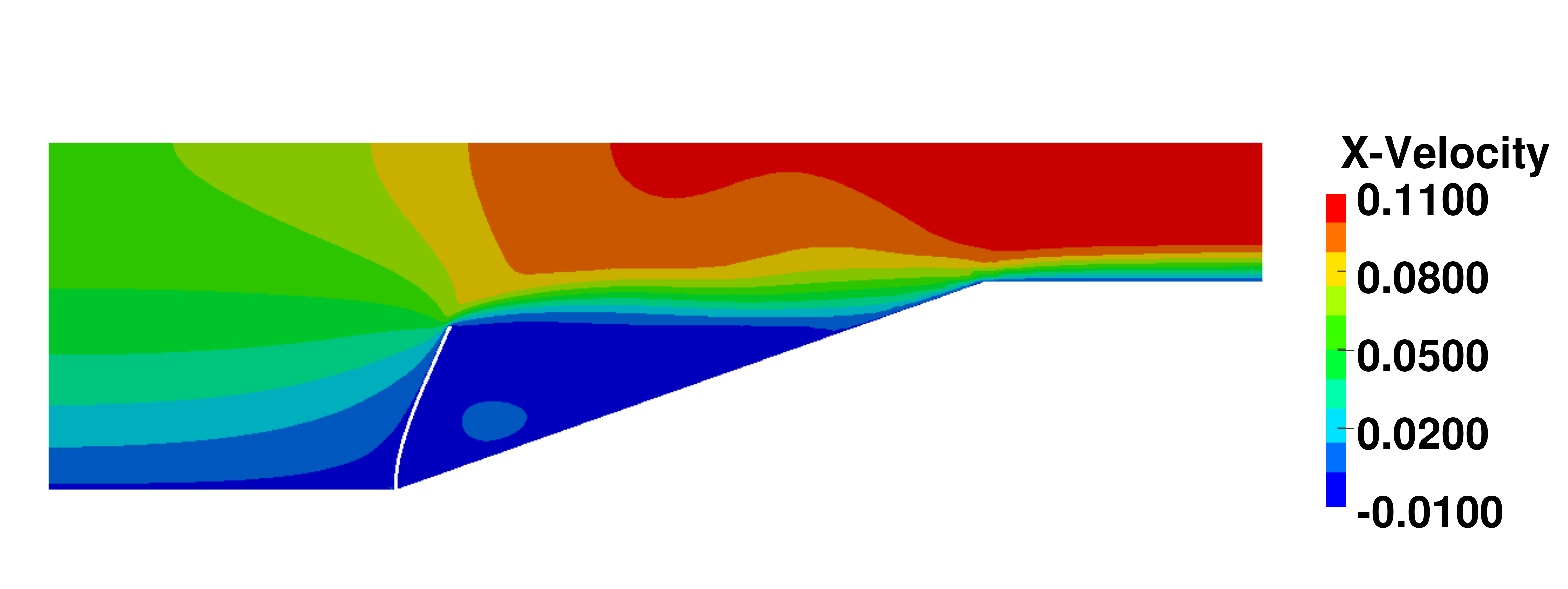}}
\subfigure[Proposed scheme]{\includegraphics[trim = 0mm 0mm 0mm 0mm, clip, scale=0.275]{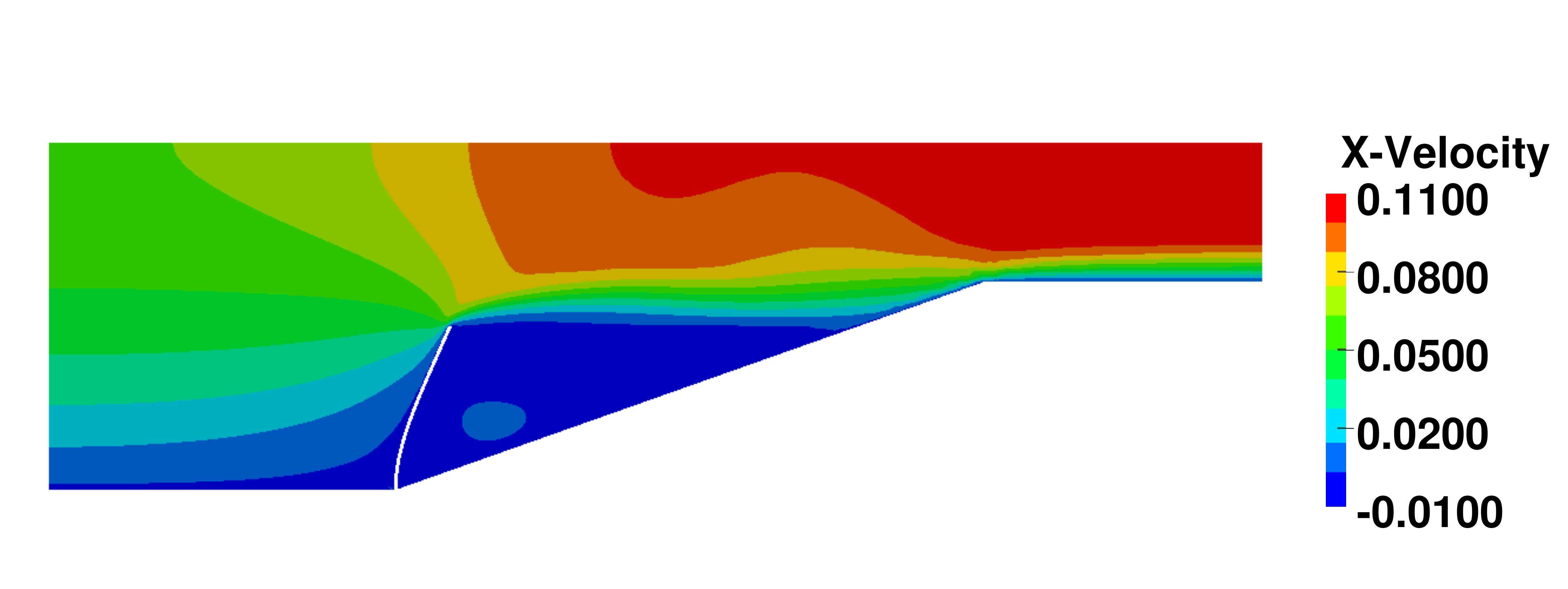}}
\caption{Flexible restrictor flap: contour plots of velocity (m/s) at $t=50$ s.}
\label{fig-flap-contours-velo}
\end{figure}

\begin{figure}[H]
\centering
\subfigure[Standard scheme]{\includegraphics[trim = 0mm 0mm 0mm 0mm, clip, scale=0.275]{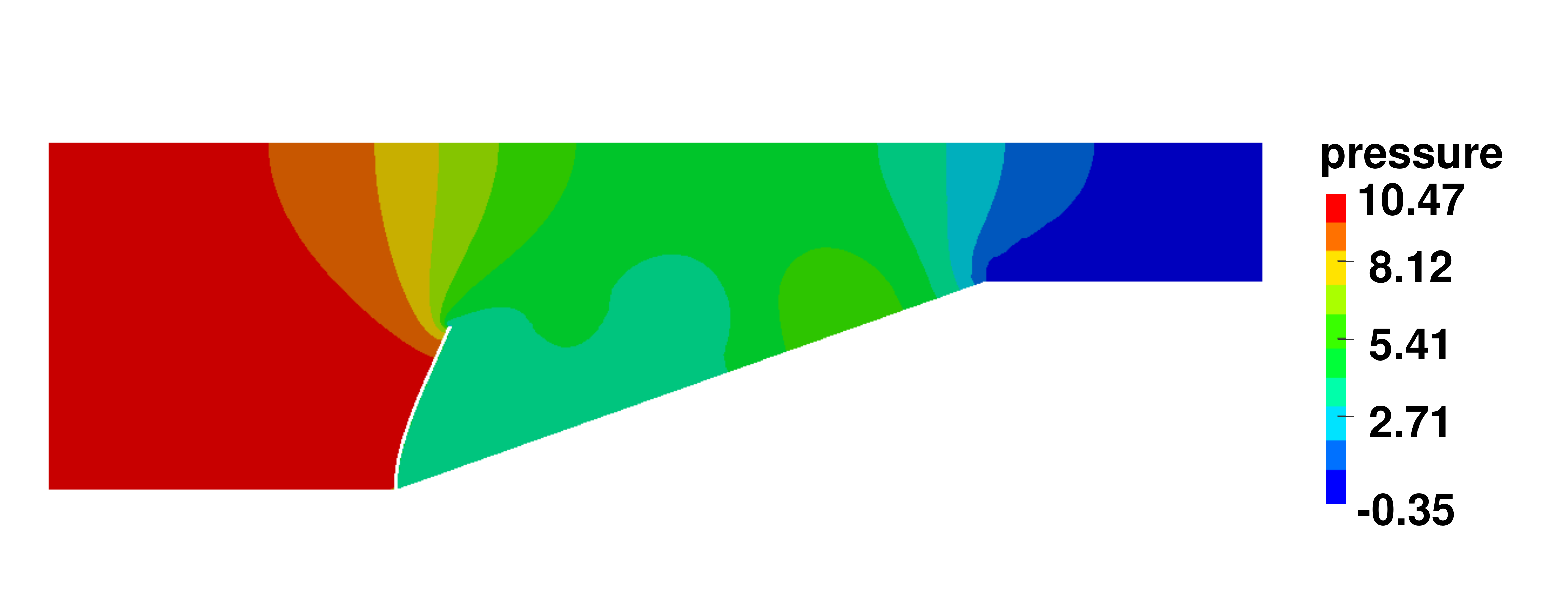}}
\subfigure[Proposed scheme]{\includegraphics[trim = 0mm 0mm 0mm 0mm, clip, scale=0.275]{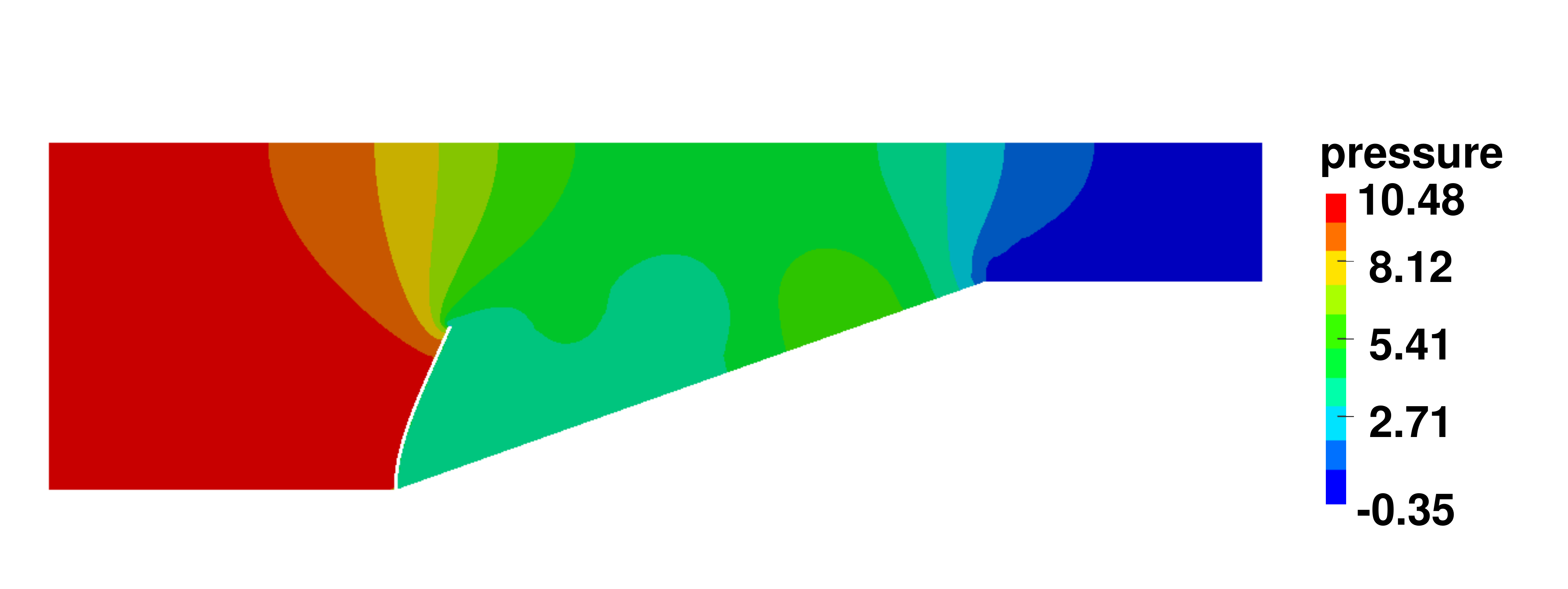}}
\caption{Flexible restrictor flap: contour plots of pressure (Pa) at $t=50$ s.}
\label{fig-flap-contours-pres}
\end{figure}

\begin{figure}[H]
\centering
\includegraphics[trim = 0mm 0mm 0mm 0mm, clip, scale=0.8]{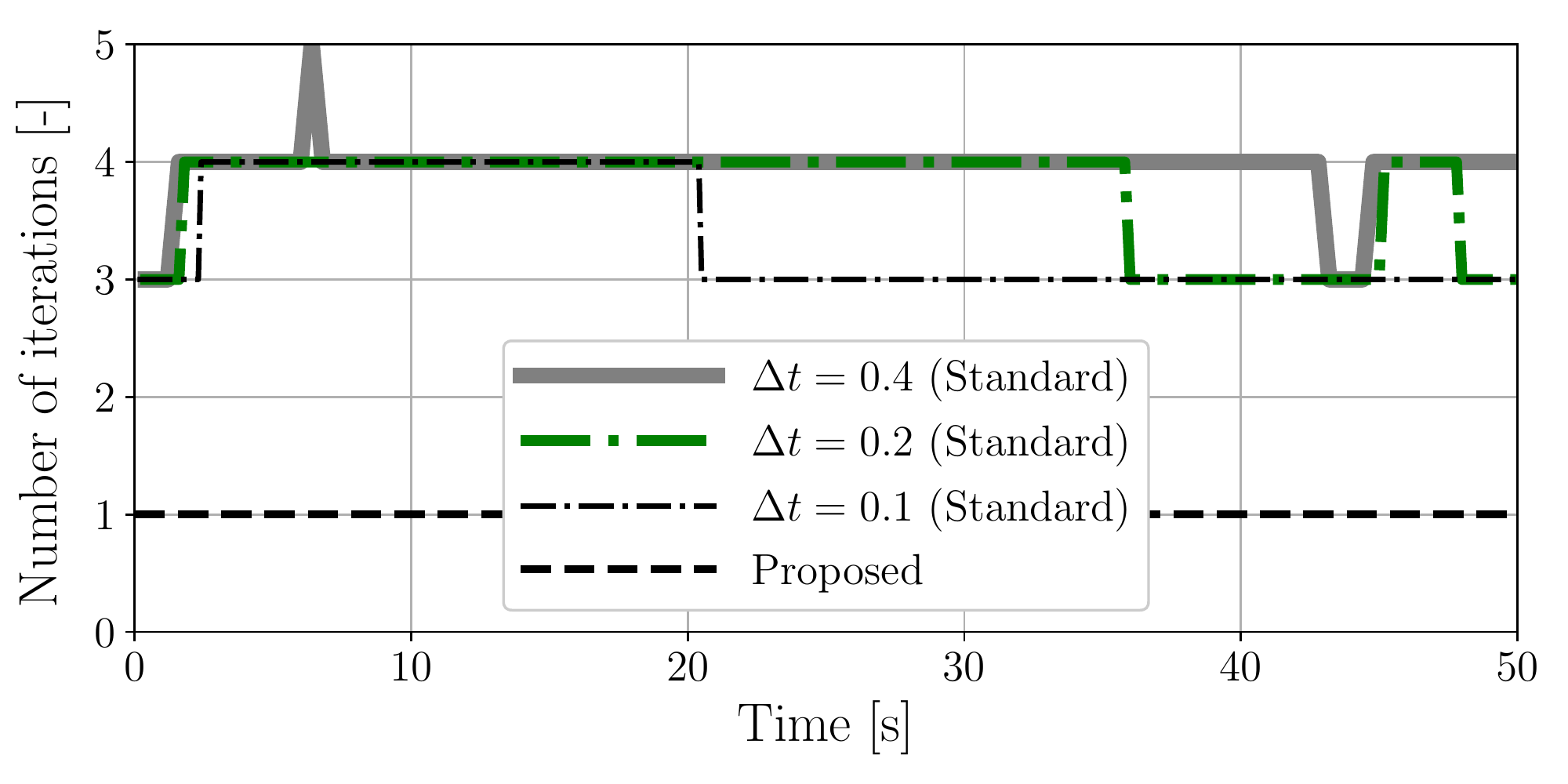}
\caption{Flexible restrictor flap: number of iterations for the fluid problem at each time step with different time step sizes. The proposed scheme requires only one iteration irrespective of the time step size.}
\label{fig-flap-graphs-iterations}
\end{figure}

Simulations are carried out for up to 50 seconds using three different time steps $\Delta t=\{0.4, 0.2, 0.1 \}$ using the standard nonlinear and the proposed linear schemes. The evolution of horizontal displacement of points A and B obtained from different simulations is presented in Fig. \ref{fig-flap-graphs-disp}. These results illustrate that the displacement response of the beam obtained with the standard and the proposed schemes are indistinguishable from each other for all the three different time steps considered. The accuracy of the results is further justified with the contour plots of X-velocity and pressure at $t=50$ s in Figs. \ref{fig-flap-contours-velo} and \ref{fig-flap-contours-pres}, respectively. 
Fig. \ref{fig-flap-graphs-iterations} shows the number of Newton-Raphson iterations for the fluid problem at each time step for all the simulations. These results prove that numerical results that are in excellent agreement with the standard approach can be obtained using the proposed scheme at a lower computational cost, about only 30\% of that of the standard scheme.

Another interesting point worth highlighting from the numerical results obtained for this example is the ability of the Dettmer and Peri\'c staggered scheme \cite{DettmerIJNME2013} in effectively overcoming the significant added-mass issue well-known in this particular example.
%
%
%
%
%

\subsection{Turek FSI benchmark in 2D}
\begin{figure}[H]
\centering
\includegraphics[trim = 0mm 0mm 0mm 0mm, clip, scale=0.9]{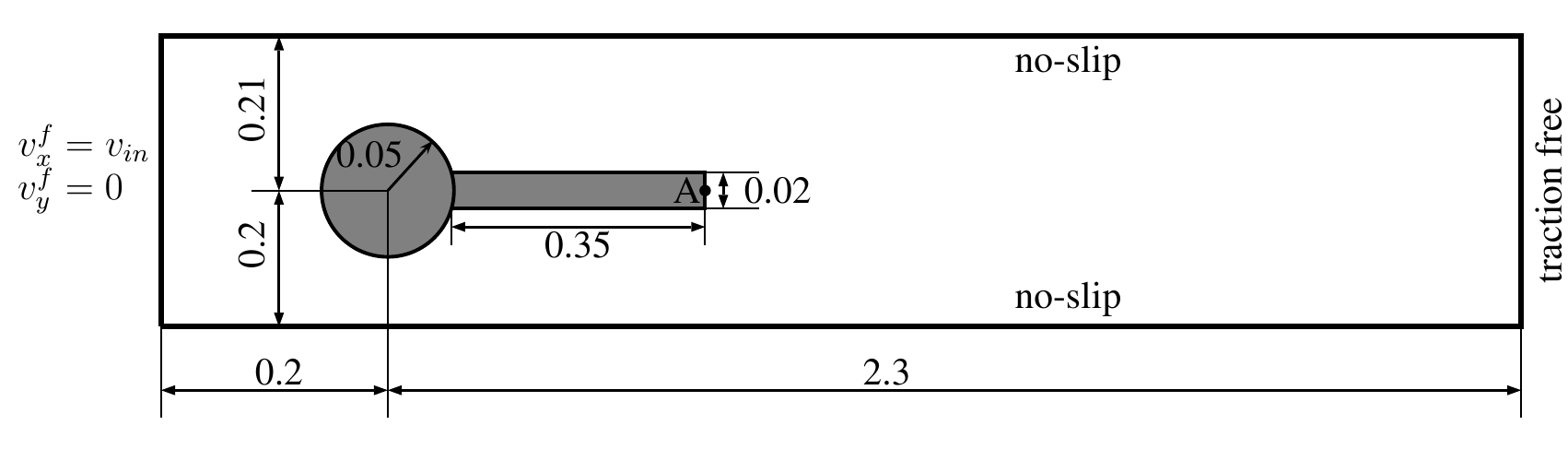}
\caption{Turek FSI2 benchmark: geometry and boundary conditions. The dimensions are in meters.}
\label{fig-turekbeam-geom}
\end{figure}

In this example, we consider the test case \textbf{FSI2} from Turek and Hron \cite{TurekFSIflex2006}. This particular case has been previously studied using the Newton-Raphson scheme for the fluid problem in \cite{KadapaCMAME2018}. The set up of the problem in terms of the geometry and boundary conditions is as shown in Fig. \ref{fig-turekbeam-geom}. The density and viscosity of the fluid are $\rho^f=10^3$ kg/m$^3$ and $\mu^f=1$ Pa s, respectively; and the density, Young's modulus and Poisson's ratio of the beam, respectively, are $\rho^s=10^4$ kg/m$^3$, $E^s=1.4\times 10^6$ N/m$^2$ and $\nu^s=0.4$.

\begin{figure}[H]
\centering
 \subfigure[Hierarchical b-spline mesh for the fluid.]{\label{turekbeam-mesh-fluid} \includegraphics[trim = 0mm 0mm 0mm 0mm, clip, scale=0.4]{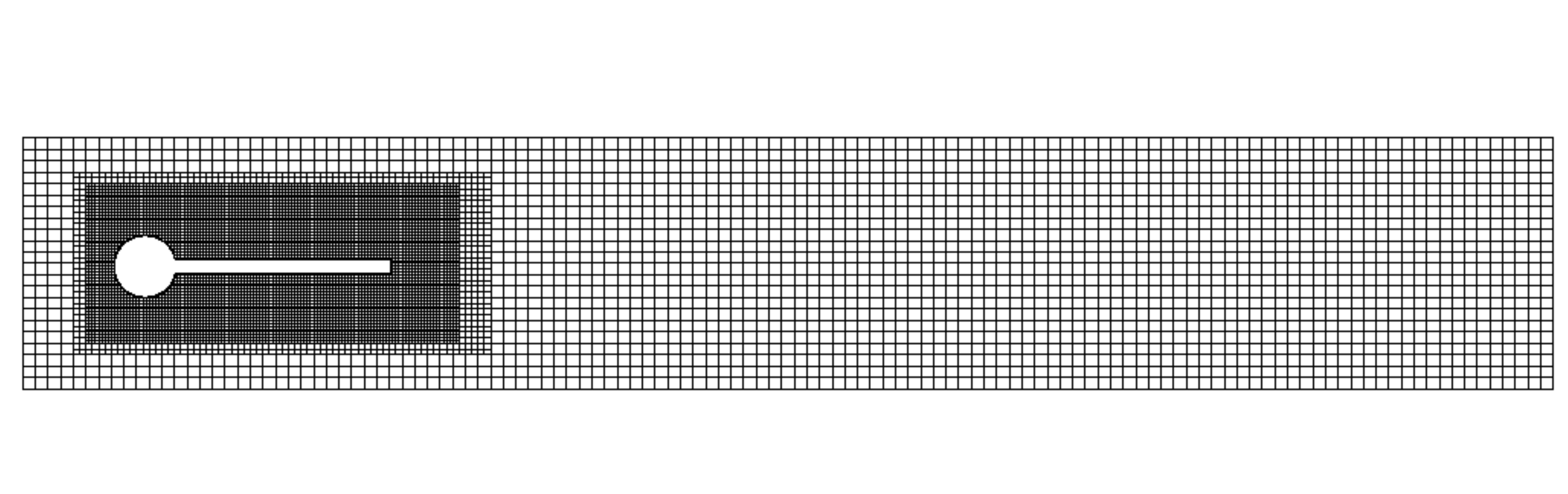}}
 \subfigure[Finite element mesh used for the beam.]{\label{turekbeam-mesh-solid} \includegraphics[trim = 0mm 0mm 0mm 0mm, clip, scale=0.3]{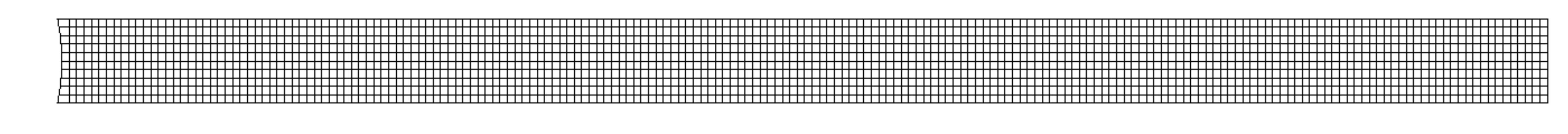}}
\caption{Turek FSI2 benchmark: meshes used for the fluid and solid domains.}
\end{figure}

The hierarchical b-spline mesh used for the fluid domain is shown in Fig. \ref{turekbeam-mesh-fluid}, and it is used with linear b-splines, resulting in approximately 26000 degrees of freedom at each time step. The beam is discretised with $200 \times 10$ bilinear quadrilateral elements with the $\boldsymbol{F}$-bar formulation of \cite{NetoIJSS1996}. The material model considered for the beam is the Saint Venant-Kirchhoff model. The relaxation parameter for the staggered scheme is set to $\beta=0.05$. The horizontal velocity profile at the inlet ($v_{in}$) is taken as,
\renewcommand{\arraystretch}{1.5}
\begin{equation}
 v_{in} = \left \{
 \begin{array}{ll}
      0.5 \, v^{y}_{in} \, [1 - \cos(\pi \, t)], & \textrm{if} \quad 0 \leqslant t \leqslant 1 \\
      v^{y}_{in}, & \textrm{if} \quad t \geqslant 1.
    \end{array}
  \right.
\end{equation}
\renewcommand{\arraystretch}{1.0}
where
\begin{equation}
v^{y}_{in} = \frac{6}{0.1681} \, y \, [0.41-y] \, \textrm{ m/s}.
\end{equation}

\begin{figure}[H]
\begin{center}
 \includegraphics[trim = 0mm 0mm 0mm 0mm, clip, scale=0.6]{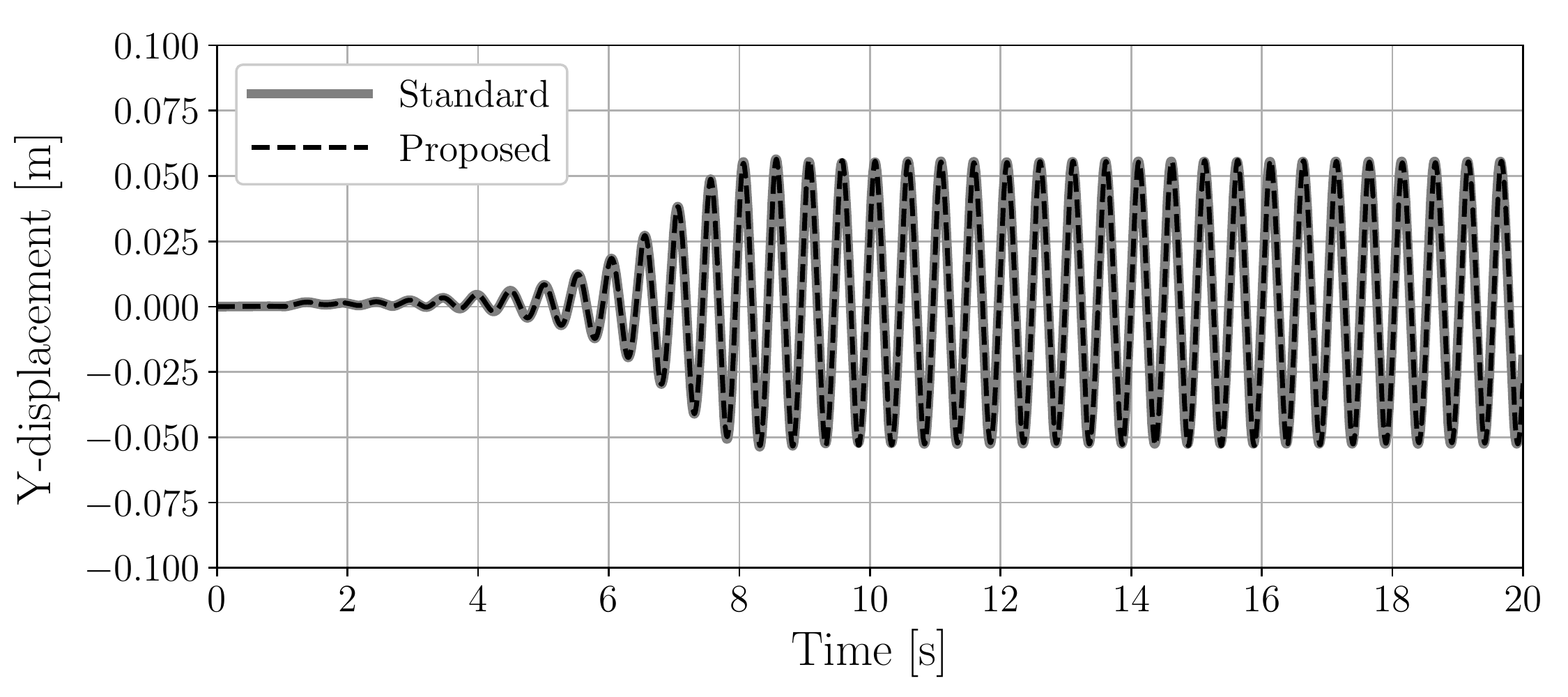}
\caption{Turek FSI2 benchmark: Y-displacement of point A obtained with $\Delta t = 0.008$.}
\label{fig-turekbeam-graph-disp-dt0p008}
\end{center}
\end{figure}

\begin{figure}[H]
\begin{center}
 \includegraphics[trim = 0mm 0mm 0mm 0mm, clip, scale=0.6]{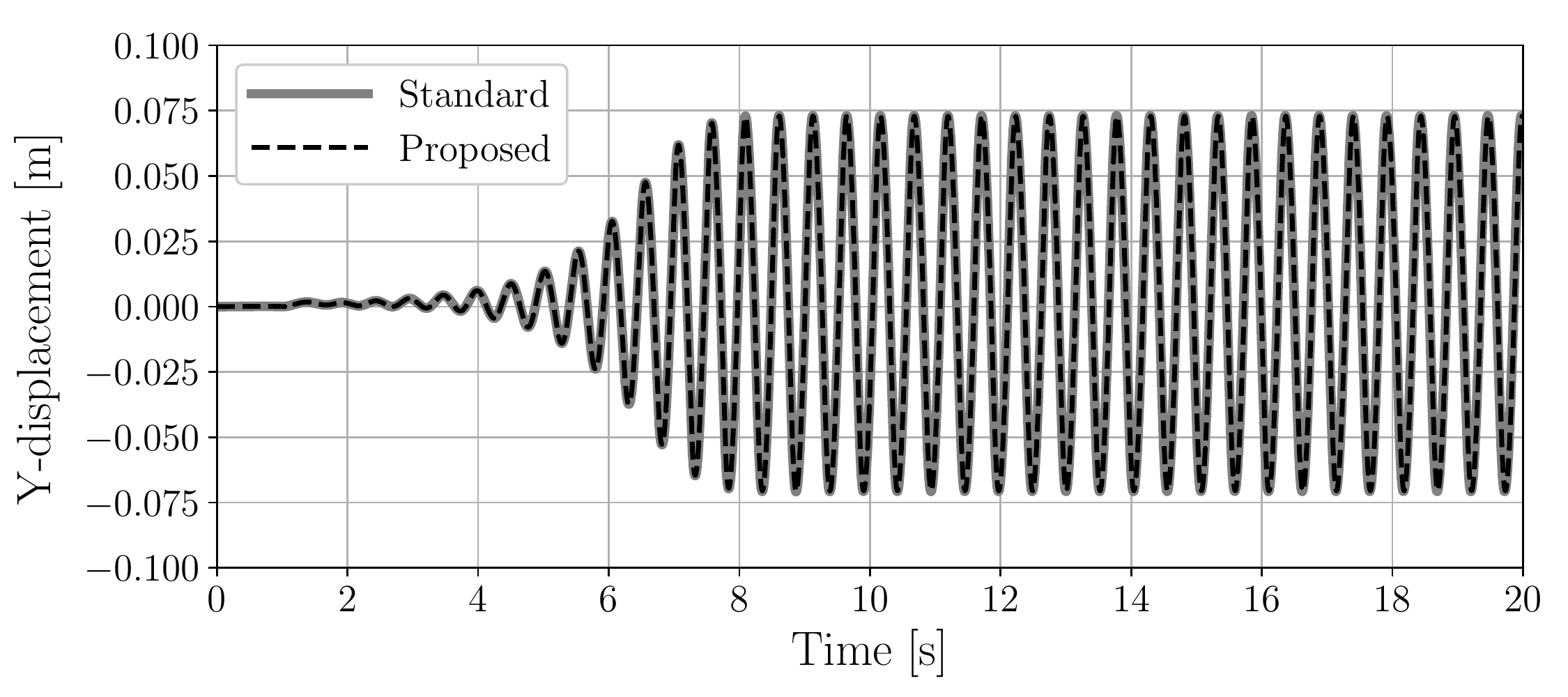}
\caption{Turek FSI2 benchmark: Y-displacement of point A obtained with $\Delta t = 0.004$.}
\label{fig-turekbeam-graph-disp-dt0p004}
\end{center}
\end{figure}

\begin{figure}[H]
\begin{center}
 \includegraphics[trim = 0mm 0mm 0mm 0mm, clip, scale=0.6]{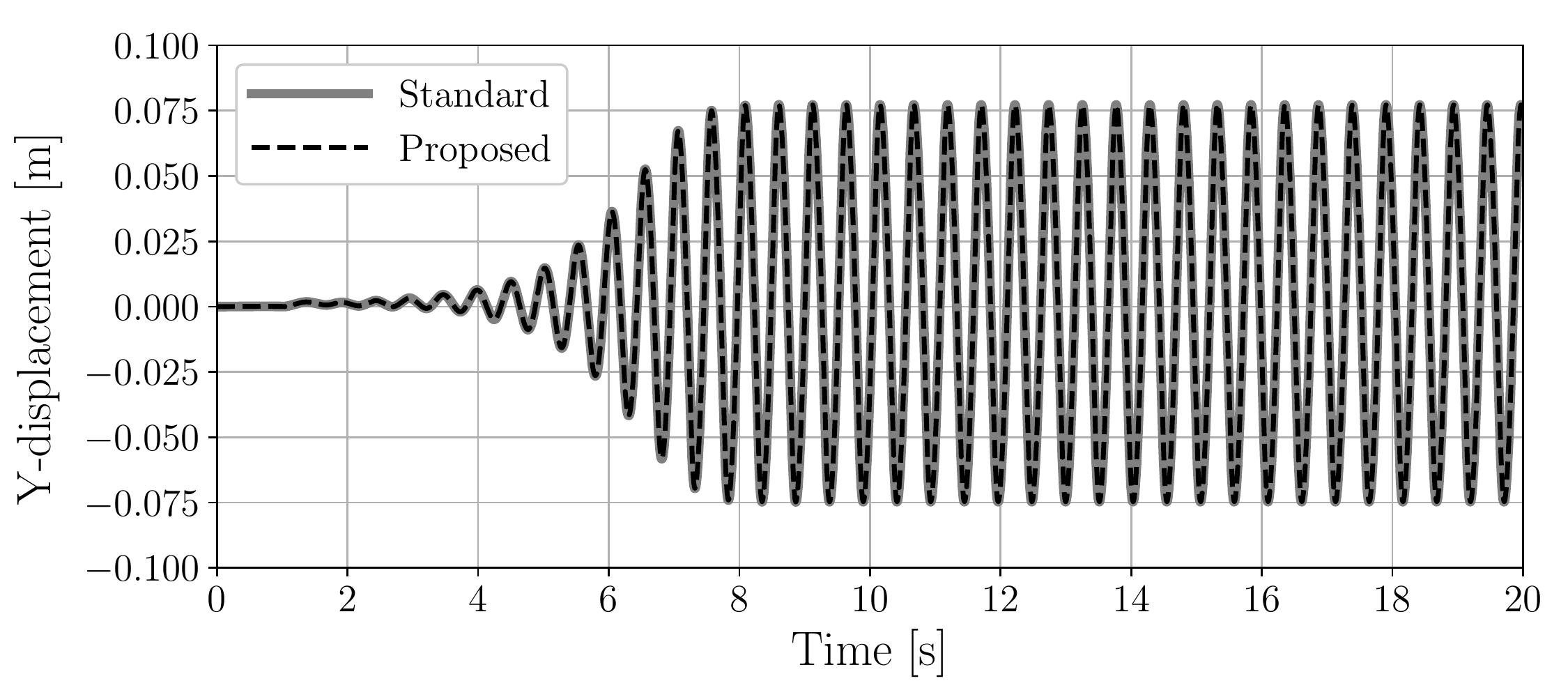}
\caption{Turek FSI2 benchmark: Y-displacement of point A obtained with $\Delta t = 0.002$.}
\label{fig-turekbeam-graph-disp-dt0p002}
\end{center}
\end{figure}

\begin{figure}[H]
 \begin{center}
 \includegraphics[trim = 0mm 0mm 0mm 0mm, clip, scale=0.7]{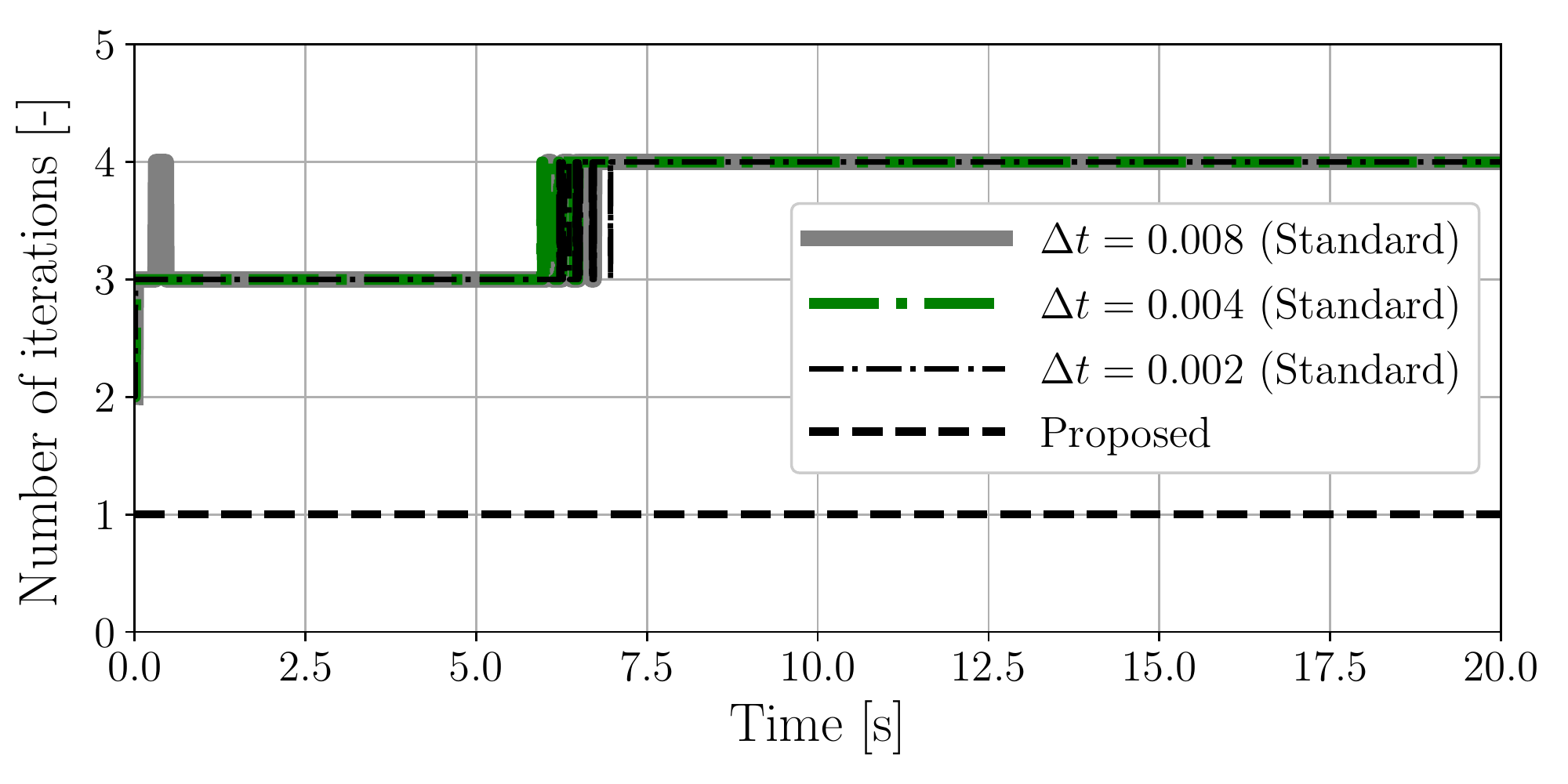}
 \caption{Turek FSI2 benchmark: number of iterations for the fluid problem at each time step with different time step sizes.}
\label{fig-turekbeam-graphs-iterations}
 \end{center}
\end{figure}

To assess the relative performance of the proposed approach over the standard approach based on the Newton-Raphson scheme for the fluid problem, simulations are carried out with three different time steps, $\Delta t=\{0.008, 0.004, 0.002\}$ s. The evolution of displacement of point A in the Y-direction obtained with the standard and the proposed schemes are presented in Figs. \ref{fig-turekbeam-graph-disp-dt0p008}, \ref{fig-turekbeam-graph-disp-dt0p004} and \ref{fig-turekbeam-graph-disp-dt0p002}, respectively, for $\Delta t=0.008$, $\Delta t=0.004$ and $\Delta t=0.002$. The maximum displacement and frequency of oscillations are tabulated in Table. \ref{table-turekbeam}. From these results, it is clearly evident that the results obtained with the proposed scheme are in excellent agreement with those of the standard scheme. The number of iterations for the fluid problem, as shown in Fig. \ref{fig-turekbeam-graphs-iterations}, illustrate that the proposed scheme is at least three times faster than the standard scheme. Thus, accurate numerical results can be obtained with the proposed iteration-free scheme using fewer computational resources when compared with the standard technique based on the Newton-Raphson scheme. This computational advantage of the proposed scheme are directly transferable to large-scale three-dimensional problems.
%
%
%

\renewcommand{\arraystretch}{1.5}
\begin{table}[H]
\centering
\begin{tabular}{|p{8cm}|>{\centering}p{4cm}|>{\centering\arraybackslash}p{2cm}|}
\hline
  Data       &  max($d^s_y \times 10^3$) & $f_o$ \\
\hline
Turek and Hron \cite{TurekFSIflex2006} - Level-2, $\Delta t=0.001$  &   $1.18 \pm 78.80$   &  2.00  \\
Turek and Hron \cite{TurekFSIflex2006} - Level-3, $\Delta t=0.001$  &   $1.25 \pm 79.30$   &  2.00  \\
Turek and Hron \cite{TurekFSIflex2006} - Level-4, $\Delta t=0.001$  &   $1.23 \pm 80.60$   &  2.00  \\
\hline
Standard scheme   \; -- \; $\Delta t=0.008$     &   $1.40 \pm 55.00$   &  2.00  \\
Proposed scheme   \; -- \; $\Delta t=0.008$     &   $1.50 \pm 54.90$   &  2.00  \\
\hline
Standard scheme   \; -- \; $\Delta t=0.004$     &   $1.15 \pm 71.95$   &  1.95  \\
Proposed scheme   \; -- \; $\Delta t=0.004$     &   $1.15 \pm 71.75$   &  1.95  \\
\hline
Standard scheme   \; -- \; $\Delta t=0.002$     &   $1.20 \pm 76.00$   &  1.95  \\
Proposed scheme   \; -- \; $\Delta t=0.002$     &   $1.20 \pm 76.00$   &  1.95  \\
\hline
\end{tabular}
\caption{Turek FSI2 benchmark: summary of vertical displacement of point A ($d^s_y$) and frequency of oscillations ($f_o$).}
 \label{table-turekbeam}
\end{table}
\renewcommand{\arraystretch}{1.0}
\section{Summary and conclusions}  \label{section-conclusion}
We have presented a novel iteration-free approach for computing accurate numerical solutions of laminar incompressible Navier-Stokes using the mixed velocity-pressure formulation. The accuracy and the computational advantages of the proposed technique are demonstrated using four numerical examples.

The proposed scheme is proven to converge with second-order accuracies in time for both the primary variables, velocity and pressure, using an example with a manufactured solution. Later, with the example of flow past a fixed circular cylinder, the accuracy and computational benefits of the proposed scheme for unsteady laminar flow problems over fixed geometries are demonstrated. Finally, the computational advantages of the proposed scheme for simulating the multiphysics problem of fluid-structure interaction are showcased.

Important features of the proposed scheme are summarised as follows.
\begin{enumerate}
\item \emph{The proposed scheme is linear}. Therefore, it does not require an iterative technique for solving it, thereby resulting in significant savings in computational time. The resulting computational gains increase with the increase in the value of the Reynolds number and/or increasing the time step size, as illustrated with the example of flow past a circular cylinder.
\item For large steps, numerical solutions computed using the proposed iteration-free scheme are in excellent agreement with the standard nonlinear scheme when compared with the scheme based on the extrapolated convection velocity.
\item Significant computational benefits can be achieved only with minor modifications to the existing code, even without any parallelisation.
\item The proposed approach is not limited to the FSI framework used in the present work. The demonstrated computational benefits could also be realised by adapting the proposed approach in any staggered or partitioned FSI algorithm, either with body-fitted or unfitted meshes, that employs a fluid solver that is based on the mixed velocity-pressure formulation with or without stabilisation.
\end{enumerate}

From the results presented in this paper, we conclude that the proposed scheme is computationally appealing for computing the numerical solutions of laminar incompressible Navier-Stokes and its extension to FSI problems. It is apparent from the presented results that the computational savings resulting from the proposed scheme would be substantial, especially for large-scale fluid-structure interaction problems in three-dimensions. We do not envisage any issues in extending this scheme to problems in three-dimensions.

\section*{ACKNOWLEDGEMENTS}
The first author acknowledges the support of the Supercomputing Wales project, which is part-funded by the European Regional Development Fund (ERDF) via the Welsh Government.

\section*{References}
\bibliographystyle{unsrt}

\end{document}